\documentclass[a4paper,fleqn,usenatbib]{mnras}

\usepackage{graphicx}	
\usepackage{amsmath}	
\usepackage{amssymb}	
\usepackage{multicol}        
\usepackage{bm}		
\usepackage{pdflscape}	

\title[Testing MOG, Non-Local Gravity and MOND with rotation
  curves of dwarf galaxies]{Testing MOG, Non-Local Gravity and MOND with rotation
  curves of dwarf galaxies} \author[M.H. Zhoolideh Haghighi and
S. Rahvar]{M.H. Zhoolideh Haghighi$^{ }$
  \thanks{zhoolideh_m@sharif.edu}, S. Rahvar$^{ }$\thanks{rahvar@sharif.edu}  \\
  $^{ }$Department of Physics, Sharif University of Technology, P.O.
  Box 11155-9161, Tehran, Iran}

\pagerange{\pageref{firstpage}--\pageref{lastpage}} \pubyear{}
\def\LaTeX{L\kern-.36em\raise.3ex\hbox{a}\kern-.15em
	T\kern-.1667em\lower.7ex\hbox{E}\kern-.125emX}

\begin{document}
	\label{firstpage}
	\pagerange{\pageref{firstpage}--\pageref{lastpage}}

\maketitle

\begin{abstract}
{ Modified Gravity (MOG) and Non-Local Gravity (NLG) are two
  alternative theories to General Relativity.  They are able to
  explain the rotation curves of spiral galaxies and clusters of
  galaxies without including dark matter
  \citep{rahvar1,rahvar2,rahvar3}.  In the weak-field approximation
  these two theories have similar forms, with an effective
  gravitational potential that has two components: (i)
  \mbox{Newtonian} gravity with the gravitational constant enhanced by
  a factor $(1+\alpha)$ and (ii) a Yukawa type potential that produces a
  repulsive force with length scale $1/\mu$. In this work we compare
  the rotation curves of dwarf galaxies in the LITTLE THINGS catalog
  with predictions of MOG, NLG and Modified Newtonian Dynamics
  (MOND). We find that the universal parameters of these theories, can fit the rotation curve 
  of dwarf galaxies with a larger stellar mass to the light ratio compared to the nearby stars in the Milky 
  Way galaxy. Future direct observations of mass function of stars in the dwarf galaxies can examine different modified gravity models.}
\end{abstract}

\section{Introduction}
At the scales of galaxies and clusters of galaxies, observations show
a systemic discrepancy between dynamical mass models and the mass
distributions inferred from the luminous
matter~\citep{zw,rubin1,rubin2}.  One proposal for resolving this
discrepancy is dark matter---so-called missing mass of the
Universe.  Cosmological dark matter is a fluid composed of massive
particles that interact gravitationally with each other, with the
possibility of very weak non-gravitational interaction with themselves
and with ordinary (baryonic) matter. {The most accurate and acceptable model of dark matter is $\Lambda CDM$ which by having six parameters explains CMB data~\citep{CDM}, large scale structure of the universe~\citep{StFormation} and Baryonic Acoustic Oscillations \citep{BAO}. It should be noted that it is hard for any alternative theory to explain such a wide range of observations that $\Lambda CDM$ do. While dark matter 
is successful in interpretation of observations, no explicit signal of dark
matter particle interaction with the ordinary matter has yet been found
\citep{moore,gal,ang,ake}.} {Recent observations of 153 galaxies with different morphology, mass, size and gas fraction shows that there is a strong correlation between observed radial acceleration and  acceleration results from the baryonic matter ~\citep{McGaugh}, which maybe suggests new dynamical laws rather than dark matter. Moreover in the gravitational lensing, \cite{Sanders} showed that mass derived from the lensing within the Einstein ring has linear correlation with the surface brightness.}

An alternative approach to interpret the dynamics of large structures is to analyze observations using a modified
law of gravity, with no dark matter.  One well known model is Modified
Newtonian Dynamics (MOND), which changes the Newtonian dynamics at
small accelerations in a way that produces flat rotation curves for
spiral galaxies \citep{milgrom}. MOND was extended to a relativistic
theory by \citet{beken}.  Some challenges facing MOND and other
modified gravity theories are to explain the gravitational lensing of systems like the bullet cluster
\citep{bullet} and the large scale structure formation in the
Universe without using dark matter. { Although it seems that MOND can not be made consistent with the detailed
shape of the CMB and matter power spectra,  there are some works on hybrid models which include both DM and MOND phenomena \citep{hybrid1,hybrid2} and explain both the dynamic of galaxies and cosmological observations.}

Another modification to the gravity is done by \citet{mashhoon} where they introduced Non-Local Gravity (NLG), with a modified
gravitational acceleration, to solve the missing mass problem.  NLG
extends non-local special relativity to the accelerating frames. In the
weak-field, non-relativistic limit of NLG, the effective gravitational
potential of a point mass adds to a Newton-like potential a new term
that gives a repulsive, Yukawa-like force \citep{mashhoon}.
Predictions of this theory have been compared with the rotation curves
of spiral galaxies and the temperature profiles of hot gas for
clusters of galaxies in the Chandra database.  With the fixed values for
the parameters of NLG, the dynamics of spiral galaxies and clusters of
galaxies are consistent with the baryonic distribution of matter in
these systems, with no need for dark matter \citep{rahvar3}.

Modified Gravity (MOG) theory---a covariant extension of General
Relativity---also avoids the need for dark matter \citep{moffat06}.
In MOG, gravity is described by the tensor metric field in combination
with new scalar and vector fields.  An important feature is that each
particle has a fifth force charge, proportional to its inertial mass,
through which it couples to the massive vector field.  Similar to the
Lorentz acceleration of charged particles in electrodynamics, test
particles in MOG deviate from geodesics due to coupling of their fifth
force charge with the vector field.  In the weak-field approximation a
modified Poisson equation is obtained, which for a point-like mass has
a Yukawa repulsive term in addition to a conventional Newtonian
potential.  Comparison with the dynamics of spiral galaxies in the
THINGS catalog results in universal values for the parameters of this
model, with reasonable fits to the rotation curves of galaxies and
clusters of galaxies \citep{rahvar1,rahvar2}. { For all the modified gravity models, the crucial observational tests would be predicting (i) the angular power spectrum of CMB (ii) the power spectrum of large scale structures and the other consequences of structure formation as the Baryonic acoustic oscillations. }

Although NLG and MOG
have completely different physical axioms, they have almost same behaviour in the weak-field
approximation limit. To test the universality of parameters of modified gravity models at
intermediate scales, 
we use the dynamical and luminosity data of dwarf galaxies in the LITTLE THINGS catalog
and interpret them in the effective potentials of MOG, NLG, and MOND.
The observational data in the LITTLE THINGS catalog are the density
distributions of stars and gas of each galaxy as well as the dynamics
in form of galaxy rotation curves.


In Section \ref{mogfield}, we review three alternative theories of
gravity of  (i) Modified Gravity (MOG) and its weak field
approximation (ii) Non-Local Gravity (NLG) and the corresponding weak field
approximation (iii) and Modified Newtonian Dynamics (MOND).  In Section
\ref{rotcurve}, we introduce the LITTLE THINGS catalog and apply the
results of weak field approximation of alternative models of gravity
to the dwarf galaxies.  We determine the best-fitting values for the
stellar mass to light ratio for our three model. {
The conclusion is given in Section \ref{conc} where we discuss the universality of
parameters of modified gravity theories and a larger stellar mass to light ratio for dwarf galaxies 
which can be used to examine the modified gravity models with the future direct observations
 of stellar mass function in these galaxies.}

\section{ Alternative theories to dark matter}
\label{mogfield} 
One of the important questions in cosmology is how to interpret the
dynamics of structures such as clusters of galaxies and spiral
galaxies, for which the observed baryonic matter is not enough
\citep{zw,rubin1,rubin2}.  An alternative to assuming the existence of
dark matter is to modify the theory of gravity in a way that can
explain the observations with only baryonic matter.  In this section
we introduce three models for the modification of gravity law.
Modified Gravity (MOG) which was introduced by \citet{moffat06} and
the weak field approximation detailed in \citet{rahvar1}.  Non-Local
Gravity (NLG), where the Einstein equations 
in the teleparallel form are similar to the Maxwell equations, hence these equations can be written in a non-local way \citep{mashhoon}. MOdified
Newtonian Dynamics (MOND) modifies Newton's law of gravity for small
accelerations \citep{milgrom}.

\subsection{Field equations in MOG}
We use the metric signature convention $(-,+,+,+)$. The general form
of action for MOG, which is also called Scalar-Tensor-Vector-Gravity
(STVG), is given by \citep{moffat06,moffat09}:
\begin{equation}
\label{action1}
S=S_G+S_\phi+S_S+S_M,
\end{equation}
where $S_G$ is the standard Einstein action:
\begin{equation}
S_G=\frac{1}{16\pi}\int\frac{1}{G}\left({\it R}+2\Lambda\right)\sqrt{-g}~d^4x,
\end{equation}
and $S_\phi$ and $S_S$ are the actions for the massive vector and
scalar fields and take the forms:
\begin{eqnarray}
  S_\phi&=&\frac{1}{4\pi}\int\omega\Big[\frac{1}{4}{\bf\it B^{\mu\nu}B_{\mu\nu}}-
  \frac{1}{2}\mu^2\phi_\mu\phi^\mu\nonumber\\
  &+&V_\phi(\phi_\mu\phi^\mu)\Big]\sqrt{-g}~d^4x,
\end{eqnarray}
and 
\begin{eqnarray}
  S_S&=&\int\frac{1}{G}\Big[\frac{1}{2}g^{\alpha\beta}({G^{-2}}
  {\nabla_\alpha G\nabla_\beta G}
  +{\mu^{-2}}{\nabla_\alpha\mu\nabla_\beta\mu})\nonumber\\
  &-&{G^{-2}}{V_G(G)}-{\mu^{-2}}{V_\mu(\mu)}\Big]\sqrt{-g}~d^4x.
\label{scalar}
\end{eqnarray}
Here $\nabla_\nu$ is the covariant derivative with respect to the
metric $g_{\mu\nu}$; the Faraday tensor of the vector field is defined
by $B_{\mu\nu}=\partial_\mu\phi_\nu-\partial_\nu\phi_\mu$; $\omega$ is
a dimensionless coupling constant; $G$ is a scalar field representing
the gravitational coupling strength; and $\mu$ is a scalar field
corresponding to the mass of the vector field. Also
$V_\phi(\phi_\mu\phi^\mu)$, $V_G(G)$ and $V_\mu(\mu)$ are the
self-interaction potentials associated with the vector field and the
scalar fields, respectively.  In our analysis we consider a simplified
version of this action, setting all the potentials to zero and making
$\omega$ and $\mu$ constant parameters.

The action for pressureless dust can be written as
\begin{equation}
  S_M = \int( \rho \sqrt{-u^\mu u_\mu} - \omega
  J^\mu\phi_\mu)\sqrt{-g} dx^4,
\label{SM}
\end{equation}

For a point mass particle we substitute $\rho(x) = m \delta^3(x)$. Varying this action results in the geodesic equation:

\begin{equation}
\label{em5}
\frac{du^\mu}{d\tau} + \Gamma^\mu_{\alpha\beta}u^\alpha u^\beta =
\omega\kappa B^\mu{}_\alpha u^\alpha.
\end{equation}
We note that $m$ has cancelled out.  The acceleration term on the
right hand side of this equation causes deviation from geodesics
similar to the Lorentz acceleration in the electrodynamics---the only
difference is that the vector field in this action is massive, and
hence provides a short range interaction.

At astrophysical scales we use the non-relativistic, weak field
approximation of MOG.  This has been shown to yield good agreement
with the dynamics of spiral galaxies and clusters of galaxies
\citep{rahvar1}.  Here, we review the same procedure by expanding
fields around the Minkowski space-time in the action as follows:
\begin{equation}
  g_{\mu\nu} = \eta_{\mu\nu} + h_{\mu\nu},
\end{equation}
where $\eta_{\mu\nu}$ is the Minkowski metric. For the vector field,
we write
\begin{equation}
  \phi_\mu = \phi_{\mu(0)} + \phi_{\mu (1) },
\end{equation}
where $\phi_{\mu(0)}$ is the zeroth order and $\phi_{\mu(1)}$ is the
first order perturbation of the vector field.  For Minkowski
space-time, we set $\phi_{\mu(0)}$ equal to zero since in the absence
of matter there is no gravity source for the vector field $\phi_\mu$.
We also perturb the energy-momentum tensor about the Minkowski
background:
\begin{equation}
  T_{\mu\nu}=T_{\mu\nu(0)}+T_{\mu\nu(1)}, 
\end{equation}
where $T_{\mu\nu(0)}$ is zero. 


By substituting the perturbed forms in the action, varying the action
with respect to the metric, and ignoring the higher orders of
perturbation, we get the field equation
\begin{equation}
  R_{\mu\nu(1)} - \frac{1}{2} R_{(1)} \eta_{\mu\nu} =   8\pi G_0 T_{\mu\nu(1)}^{(M)}
  +8\pi G_0 T_{\mu\nu(1)}^{(\phi)}, \label{ein1}
\end{equation}
where $T_{\mu\nu(1)}^{(M)}$ represents the energy-momentum tensor of
matter, and $T_{\mu\nu(1)}^{(\phi)}$ is the energy-momentum tensor of
the vector field given by
\begin{eqnarray}
  T_{\mu\nu}^{(\phi)} &=& -\frac{\omega}{4\pi}(B_\mu{}^{\alpha}
  B_{\nu\alpha} - \frac{1}{4}
  g_{\mu\nu}B^{\alpha\beta}B_{\alpha\beta})\nonumber \\
  &+&\frac{\mu^2\omega}{4\pi}(\phi_\mu\phi_\nu -
  \frac{1}{2}\phi_\alpha\phi^\alpha g_{\mu\nu}).
\label{tphi}
\end{eqnarray}




In the weak field approximation we ignore higher order terms from
the vector field in the energy momentum tensor.  For the $(0,0)$ component
of the Ricci tensor we obtain
\begin{eqnarray}
  R_{00(1)} =-\frac{1}{2}\vec\nabla^2 h_{00}.
\end{eqnarray}
Substituting into equation (\ref{ein1}) results in
\begin{equation}
  -\frac{1}{2}\vec\nabla^2( h_{00}) = 4\pi G_0 \rho.
\label{effpois}
\end{equation}
Varying the action with respect to $\phi^\mu$ gives the field
equation:
\begin{equation}
  \nabla_\nu B^{\mu\nu} - \mu^2 \phi^\mu = -\frac{4\pi}{\omega} J^\mu.
\end{equation}
Let us assume that the current, $J^\mu$, is conserved: $\nabla_\mu
J^\mu=0$ (For constant $\kappa$, this is equivalent to conservation
of the matter current $\rho u^\mu$).  In the weak field approximation
this implies the constraint $\phi^\mu{}_{,\mu} = 0$ which, for the
static case, simplifies to
\begin{equation}
  \vec\nabla^2\phi^0 -\mu^2\phi^0 = -\frac{4\pi}{\omega}J^0.
\end{equation}
This has the solution of
\begin{equation}
  \phi^0(x) = \frac{1}{\omega}\int\frac{e^{-\mu|\vec x-\vec x'|}}
  {|\vec x-\vec x'|}J^0(\vec x')d^3x'. \label{phi0}
\end{equation}

Taking the divergence of the test particle equation of motion
(\ref{em5}), applying the weak field approximation, and ignoring time
derivatives gives
\begin{equation}
  \vec\nabla\cdot \mathbf a  - \frac12\vec\nabla^2 h_{00} = - \omega\kappa
  \vec\nabla^2\phi^0, \label{p2}
\end{equation}
where $"\mathbf a"$ represents the acceleration of the test particle.
We define an effective potential, which determines the test particle
acceleration, by $\mathbf a = -\vec\nabla\Phi_{eff}$.  Substituting
$\vec\nabla^2h_{00}$ from (\ref{effpois}) into (\ref{p2}) relates
$\Phi_{eff}$ to the distribution of matter:
\begin{equation}
  \vec\nabla\cdot(\vec\nabla\Phi_{eff} - \kappa\omega\vec\nabla\phi^0)
  = 4\pi G_0\rho. \label{po3}
\end{equation}
Inserting $\phi^0$ from (\ref{phi0}), and setting $J^0=\kappa \omega
\rho$, gives the effective potential due to a mass distribution
$\rho(\vec{x})$:
\begin{equation}
  \Phi_{eff}(\vec x) = - \int\frac{G_0 \rho(\vec x')}{|\vec x-\vec
    x'|}d^3x' + \kappa^2\int\frac{e^{-\mu|\vec x-\vec x'|}}
  {|\vec x-\vec x'|}\rho(\vec x')d^3x'. \label{potential}
\end{equation}
Setting $\kappa=\sqrt{\alpha G_N}$ and $G_0=(1+\alpha)G_N$ gives the
conventional form derived in \citet{rahvar1}
\begin{equation}
  \Phi_{eff}(\vec x) = - G_N \left[\int\frac{\rho(\vec x')}
    {|\vec x-\vec x'|}(1+\alpha
    -\alpha e^{-\mu|\vec x-\vec x'|})d^3x' \right]. \label{mogphi}
\end{equation}
Consequently, the acceleration of a test particle is given by
\begin{eqnarray}
  \mathbf{a}(\mathbf x) &=& - G
  \int\frac{\rho(\mathbf x')(\mathbf{x}-\mathbf{x'})}
  {|{\mathbf x}-{\mathbf x'}|^3}\nonumber\\
  &\times&\Big[1+\alpha 
  -\alpha e^{-\mu|{\mathbf x}-{\mathbf x'}|}(1+{\mu}|{\mathbf x}- 
  {\mathbf x'}|)\Big]~d^3x'.
\label{mogacceleration}
\end{eqnarray}
Using this theory to analyze observations of spiral galaxies and
clusters of galaxies yields the unique values, $\alpha = 8.89\pm 0.34$
and $\mu = 0.04\pm0.004~kpc^{-1}$, that fit with the spiral galaxies and cluster of 
galaxies \citep{rahvar1,rahvar2}.

\subsection{Non-Local Gravity}
\label{nonlocal} 
Lorentz covariance in special relativity is valid for inertial
observers. The fundamental assumption of transformation of physical
laws between non-accelerating observers and accelerating observers is
that, locally, these two frames are equivalent. This is called the
equivalence principle.  However, for accelerating observers who are
real observers of the physical world, Lorentz covariance might not be
valid. This extension is modelled with the introduction of
non-locality in the accelerating frames and consequently in gravity. 


It has been argued that one must in general go beyond the locality
hypothesis of standard special relativity theory and include the past
history of an accelerated observer \citep{mashhoon93}. The extension
of non-locality to general relativity is done by an averaging
procedure where a kernel acts as the weight function for the
gravitational memory of the past events.  In Non-Local Gravity, it is
assumed that deviation from locality is proportional to $\lambda/{\cal
  L}$ where $\lambda$ is the characteristic length of phenomena and
${\cal L}$ is a characteristic length determined by the acceleration
of the observer. In the case that $\lambda/{\cal L}>1$, assuming that
the locality is broken, we cannot use the equivalence principle.


Non-Local Gravity is formulated in a tetrad formalism.  The tetrad
field relates the space time metric to a local Minkowski frame through
$g_{\mu \nu}(x)=e_\mu{}^{\hat \alpha}e_\nu{}^{\hat \beta}\eta_{\alpha
  \beta}$.  Here the tetrad field has sixteen degrees of freedom.  
It is necessary to extend the Riemannian structure of space-time by
using the Weitzenb\"ock connection where $\nabla_\mu e^{\hat
  \alpha}{}_{\nu} = 0$.  
Simplifying this equation results in $\Gamma^{\lambda}{}_{\mu\nu}=
e^\lambda{}_{\hat \alpha} \partial_\mu e^{\hat \alpha}{}_\nu$.



Similar to the electromagnetic field \citep{tp1,tp2,tp3}, we can
define the field strength as follows
\begin{equation}
  C_{\mu\nu}{}^{\hat \alpha} = \partial_\mu e_\nu{}^{\hat\alpha} -
  \partial_\nu e_\mu{}^{\hat\alpha}.
\end{equation}
We define another auxiliary field, a modified torsion tensor:
\begin{eqnarray}\label{I2}
  \mathfrak{C}_{\mu \nu}{}^{\hat{\alpha}} =\frac 12\,
  C_{\mu \nu}{}^{\hat{\alpha}} -C^{\hat{\alpha}}{}_{[\mu
    \nu]}+2e_{[\mu}{}^{\hat{\alpha}}
  C_{\nu]{\hat{\beta}}}{}^{\hat{\beta}}\,.
\end{eqnarray}
By defining the tensor density 
\begin{equation}
  {\cal H}^{\mu \nu}{}_{\rho}(x) =
  \frac{\sqrt{-g(x)}}{\kappa}\mathfrak{C}^{\mu \nu}{}_{\rho}\,,
\end{equation}
with $\kappa = 8\pi G$, the Einstein equation 
can be written in the
simpler form:
\begin{equation}\label{I4}
  \partial_{[\mu} C_{\nu \rho]}{}^{\hat{\alpha}}=0\,,
 \end{equation}
\begin{equation}\label{I5}
  \partial_\nu{\cal H}^{\mu \nu}{}_{\hat{\alpha}}
  =\sqrt{-g}~(T_{\hat{\alpha}}{}^\mu + E_{\hat{\alpha}}{}^\mu),
\end{equation}
where $T_{\hat{\alpha}}{}^\mu$ is the energy-momentum tensor of matter
and $E_{\hat{\alpha}}{}^\mu$ is the energy-momentum of the tetrad
field.  Similar to the energy-momentum tensor of electromagnetic
field, the latter is traceless. The advantage of teleparallel formalism  of general relativity is that we can write the Einstein equations in the
form of Maxwell equations in non-vacuum media. Here
$C_{\mu\nu}{}^{\hat \alpha}$ plays a role similar to the
electromagnetic field tensor $F_{\mu\nu}$ (i.e. $C_{\mu\nu}{}^{\hat
  \alpha}\rightarrow F_{\mu\nu}$ ), ${\cal H}^{\mu
  \nu}{}_{\hat{\alpha}}$ plays the role of displacement tensor of
$H_{\mu\nu}$ (i.e. ${\cal H}^{\mu \nu}{}_{\hat{\alpha}}\rightarrow
H_{\mu\nu}$) and finally $\sqrt{-g}~(T_{\hat{\alpha}}{}^\mu +
E_{\hat{\alpha}}{}^\mu)$ plays the role of current in the
electromagnetism (i.e $\sqrt{-g}~(T_{\hat{\alpha}}{}^\mu +
E_{\hat{\alpha}}{}^\mu)\rightarrow j^\mu$). 

It is well known that the constitutive relation between $F_{\mu\nu}$
and $H_{\mu\nu}$ in the electrodynamics of a generic medium is
non-local.  This non-locality is expressed through a kernel in the
integration.  This idea is extended to non-local relativity in a
covariant form. In the weak field
approximation, by expanding the tetrad field around
$\delta^{\alpha}{}_{\mu}$, equation (\ref{I5}) can be written as
\citep{rahvar3} 
\begin{equation}\label{II1B}
  \frac{\partial}{\partial
    x^\sigma}\Big[\mathfrak{C}_{\mu}{}^{\sigma}{}_{\nu}(x)+
  \int{\cal K}(x, y) \mathfrak{C}_{\mu}{}^{\sigma}{}_{\nu}(y)~d^4y\Big] =
  \kappa~ T_{\mu \nu}\,.
\end{equation}
The $(0,0)$ component of this equation reduces to the modified
Poisson equation
\begin{equation}
  \nabla^2\phi = 4\pi G(\rho + \rho_D), 
\end{equation}
where $\phi$ represents the gravitational potential and is related to
the test particle acceleration by $d^2x/dt^2 = -\nabla\phi$. Here $\rho_D$ is an extra term in the Poisson
equation that resembles dark matter, but that arose entirely from the
effects of non-local gravity.  It is given by
\begin{equation}
  \rho_D(x) =  \int q(|x-y|) \rho(y) d^3y,
\end{equation}
where $q(|x-y|)$ is the kernel of non-locality. We adapt the kernel
\begin{equation}\label{II8}
  q=\frac{1}{4 \pi \lambda_0}\frac{(1+\mu r)}{r^2}e^{-\mu r}\,.
\end{equation}
where $\mu$ is a constant with dimension of inverse length, and
$\lambda_0$ is the fundamental length scale of NLG which may depend on
the size of structure.  Here we assume $\lambda_0$ is a constant
parameter.  By introducing a dimensionless parameter $\alpha$ we can
write $\lambda_0 = \frac{2}{\mu\alpha}$.

Substituting this kernel in the definition of the potential, and
integrating, we obtain the acceleration of a test particle:
\begin{eqnarray}
\mathbf{a}(\mathbf x) &=& - G
\int\frac{\rho(\mathbf x')(\mathbf{x}-\mathbf{x'})}
{|{\mathbf x}-{\mathbf x'}|^3}\nonumber\\
&\times&\Big[1+\alpha 
-\alpha e^{-\mu|{\mathbf x}-{\mathbf x'}|}(1+\frac{\mu}{2}|{\mathbf x}-{\mathbf x'}|)\Big]~d^3x'.
\label{acceleration}
\end{eqnarray}
Comparing this acceleration with equation (\ref{mogacceleration}) for
MOG, the only difference between these two theories is an extra factor
of $1/2$ in the second term.

A more general situation can be supposed if we use one of the
following kernels:
\begin{equation}\label{II6}
  q_1=\frac{1}{4\pi \lambda_0}~ \frac{1+\mu (a_0+r)}{(a_0+r)^2}~e^{-\mu r}\,,
\end{equation}
\begin{equation}\label{II7}
  q_2=\frac{1}{4\pi \lambda_0}~ \frac{1+\mu (a_0+r)}{r(a_0+r)}~e^{-\mu r}\,,
\end{equation}
which lead to the following acceleration law for non-local gravity:
\begin{eqnarray}
  &&\mathbf{a}(\mathbf x) = - G
  \int\frac{\rho(\mathbf x')(\mathbf{x}-\mathbf{x'})}
  {|{\mathbf x}-{\mathbf x'}|^3}\nonumber\\
  &\times&\Big[1-{\cal E}(r)+\alpha 
  -\alpha e^{-\mu|{\mathbf x}-{\mathbf
      x'}|}(1+\frac{\mu}{2}|{\mathbf x}-
  {\mathbf x'}|)\Big]~d^3x'. \nonumber
	\label{acceleration2}
\end{eqnarray}
Here ${\cal E}(r)$ is either ${\cal E}_1(r)$ or ${\cal E}_2(r)$
associated, respectively, to $q_1$ and $q_2$ and given by
\begin{equation}\label{A3}
  {\cal E}_1(r)=\frac{a_0}{\lambda_0}\left \{-\frac{r}{r+a_0}e^{-\mu r}+2e^{\mu a_0} \Big[E_1(\mu a_0)-E_1(\mu a_0+\mu r) \Big] \right \}\,
\end{equation}
and
\begin{equation}\label{A4}
  {\cal E}_2(r)=\frac{a_0}{\lambda_0}e^{\mu a_0}\Big[E_1(\mu a_0)-
  E_1(\mu a_0+\mu r)\Big]\,,
\end{equation}
where $E_1(u)$ is the exponential integral function:
\begin{equation}\label{II21}
  E_1(u):=\int_{u}^{\infty}\frac{e^{-t}}{t}dt\,.
\end{equation} 
 
  
  
  
The best fitting parameter values for the simple kernel in equation
(\ref{II8}), obtained by comparing the dynamics of spiral galaxies
with the theory, are $\alpha=10.94 \pm 2.56$ and $\mu=0.059 \pm 0.028
kpc^{-1}$ \citep{rahvar3}.
  
\subsection{Modified Newtonian Dynamics (MOND)}
Another interesting alternative for addressing the missing mass
problem is the Modified Newtonian Dynamics, proposed by \citet{milgrom}.
To explain the flat rotation curves of spiral galaxies, Milgrom
assumed that Newtonian dynamics should be modified in a way that
preserves Newtonian gravity at large accelerations but is modified at
low accelerations. This theory introduces a universal acceleration
$a_{0}=1.0\times10^{-10}\,\mathrm{m\,s}^{-2}$ below which the Newtonian
dynamics is modified.  An equivalent version of this theory is the
modified gravitational acceleration $g$, sourced by a point mass $M$,
as follows:
\begin{equation}
  g=\Bigg \{
\begin{tabular}{c}    
  $GM/r^2 ~~~~  g \gg a_{0}$,   \\
  \\
  $	\sqrt{GM a_{0}}/r ~~~~  g\ll a_{0}$.
\end{tabular}
\end{equation}

To obtain a smooth transition between the two regimes, between the
Newtonian gravity and Mondian gravity, Milgrom's law is written in the
following form:
\begin{equation}
  g_{N}=\mu(\frac{g}{a_{0}})g,
\label{MOND}
\end{equation} 
where the interpolating function $\mu$ satisfies: $\mu(x)\rightarrow
1$ for $x \gg 1$ and $\mu(x)\rightarrow x$ for $x\ll 1$. A standard
choice for the interpolating function is:
\begin{equation}
  \mu(x)=x(1+x^2)^{-1/2}.
\label{mu}
\end{equation}
From this interpolating function, it is straight forward to find the
acceleration of a test particle, relative to the Newtonian
acceleration, by using equations (\ref{MOND}) and (\ref{mu}):
\begin{equation}
  g=\frac{g_{N}}{\sqrt{2}}\sqrt{1+\sqrt{1+(\frac{2a_{0}}{g_{N}})^2}}.
\label{Gmond}
\end{equation}

{It has been known for years that MOND does not completely explain the mass discrepancy in galaxy clusters \citep{MONDcluster}. However by using MOND the amount of estimated missing mass in clusters reduces significantly and roughly the unseen mass would be as much as observed baryonic mass. One candidate for this undetected mass is cosmological neutrinos with the mass of the order of 2$eV$ \citep{MOND-cluster1,MOND-cluster2}. }

\section{Rotation curves of galaxies}
\label{rotcurve} 
In this section, we use so-called dwarf galaxies to test the three
alternative gravity theories.  Our aim is to explain the dynamics of
structures, from the smallest galaxies to clusters of galaxies, without
using the concept of dark matter.  Dwarf galaxies are smaller than
spiral galaxies.  In dark matter models, some dwarf galaxies have a
higher ratio of dark matter to baryonic matter compared to spiral
galaxies \citep{DM_in_dwarf}.

Similar to the procedures used in \cite{rahvar1,rahvar2}, we use the
distribution of baryonic matter of dwarf galaxies, including stars and
interstellar baryonic gas, to analyze the galaxy rotation curves. We
assume, for simplicity, that galaxies have cylindrical symmetry. For
both MOG and NLG the radial component of acceleration can be
calculated by discretizing space into small elements and linear summing of the
accelerations due to each element as follows:
\begin{eqnarray}
  && a_r(r) = G_N\sum_{r'=~0}^\infty \sum_{\theta'=~
    0}^{2\pi}\frac{\Sigma(r')}{|r-r'|^3}(-r +
  r'cos\theta')(1+\alpha \nonumber\\
  & &-\alpha e^{-\mu|r-r'|} -\mu\alpha\beta |r-r'|e^{-\mu|r-r'|} )r'\Delta
  r'\Delta\theta',
\end{eqnarray}
where $\beta$ for MOG is $1$ and for NLG is $1/2$, and $\Sigma(r)$
represents the column density of the galaxy.
In the case of MOND, since the theory is non-linear, we first find the
Newtonian acceleration (i.e. $g_{N}$), as in standard Newtonian
gravity, and then obtain the MOND acceleration according to equation
(\ref{Gmond}).

From the observations, we have the surface Luminosity of stars, as well as the column density of gas. Here we choose a
sub-sample of nearby galaxies from the LITTLE THINGS catalog with high
resolution measurements \citep{dwarf}. Details about the observational parameters of
these galaxies are given below.






\subsection{LITTLE THINGS catalog}

Local Irregulars That Trace Luminosity Extremes, The HI Nearby Galaxy
Survey (LITTLE THINGS) is a high-resolution ($\sim 6^{''}$\,angular;
$<$2.6~km\,s${}^{-1}$ velocity resolution) survey of nearby (within
10.3 Mpc) dwarf galaxies in the local volume observed with the Very
Large Array (VLA).  Galaxies with distances $>$10~Mpc have been
excluded in order to achieve a reasonably small spatial resolution in
VLA HI maps. The mean distance to the galaxies in the \mbox{LITTLE} THINGS
sample is 3.7~Mpc. In addition galaxies with
$W_{20}>160$~km\,s${}^{-1}$ have been excluded in order to fit the
galaxy emission comfortably in the bandwidth available for the desired
velocity resolution \citep{Massmodel}. Here, $W_{20}$ is the
full-width at 20\% of the peak of an integrated HI flux-velocity
profile.

The high-resolution HI observations enable us to derive reliable
rotation curves of the sample galaxies in a homogeneous and consistent
manner. The rotation curves are then combined with the Spitzer
archival $3.6\,\mu$m and ancillary optical U, B, and V images to
construct mass models of the galaxies. Spitzer IRAC $3.6\,\mu$m images
are less affected by dust and less sensitive to the young stellar
populations compared with optical images, making them better for
tracing old stellar populations found in dwarf galaxies. This high
quality, multi-wavelength dataset significantly reduces observational
uncertainties and thus allows us to examine in detail the mass
distribution in the galaxies \citep{dwarf}.  In high resolution HI
data, observational systematic effects are properly reduced, which provides us more accurate
data.
 
The LITTLE THINGS catalog contains 37 dwarf galaxies, but the rotation curve and the mass
model of 26 dwarf galaxies are available \citep{dwarf}.  For our analysis we choose a
sub-sample of 16 dwarf galaxies for which the data has good coverage
of rotation curves.  Table (\ref{tab1}) shows
the list of galaxies in the sub-sample of the LITTLE THINGS catalog.

\subsection{Comparing modified gravity models with the observations of Little
  Things galaxies}

In this section we fit the three modified gravity models to the
dynamics of dwarf galaxies in LITTLE THINGS catalog. 
We note that In the Newtonian gravity, one of crucial parameters in the estimation of
dark matter content of a structure is the overall mass to the light
ratio of a structure where we can extract the mass from the dynamics of
structure and luminosity from the apparent luminosity of structure
after correcting for absorption in the intergalactic and galactic
media.

The luminosity of galaxies in our study is given in terms of
magnitude in V-band \citep{Lthings} and dynamics of galaxies are also
obtained from the rotation curve of galaxies \citep{dwarf}.  





\subsubsection{Fitting observations with MOG}

{
For the mass models of dwarf galaxies, we define a linear interpolation function to construct the mass
profile from the centre to the edge of galaxy. In the observation we have the brightness of 
galaxies which has to be translated to the mass profile. 
The observations for each galaxy give
data points $(x_{i},F_{i})$, where $x_i$ is the distance from the
centre of the galaxy and $F_i$ is the surface brightness at $x_i$.
We define a continuous surface brightness profile $F(x)$ for stars by setting
\begin{equation}
  F(x)=F_{i}+\frac{F_{i+1}-F_{i}}{x_{i+1}-x_{i}}(x-x_{i}),
\end{equation}
And continuous column density profile for gas by setting
\begin{equation}
  \Sigma^{gas}(x)=\Sigma_{i}^{gas}+\frac{\Sigma_{i+1}^{gas}-\Sigma_{i}^{gas}}{x_{i+1}-x_{i}}(x-x_{i}),
\end{equation}
for $x$ between $x_{i}$ and $ x_{i+1}$.
The column density of each galaxy is given by $\Sigma=\Upsilon_{*}^{(3.6)}F+\Sigma^{gas}$. Using this continuum column density, and assuming universal values for the parameters $\mu$ and $\alpha$, we can numerically compute the integral
in equations (\ref{mogacceleration}) and (\ref{acceleration}) for a test particle at radial position of
$"x"$. Moreover we do the same procedure using a universal value of $a_{0}=1.0\times10^{-10}\,\mathrm{m\,s}^{-2}$ in MOND and compute the equation (\ref{Gmond}). The radial acceleration provides the rotation curve of stars and gas, and 
consequently the overall rotational curve of the galaxies in MOG, NLG and MOND.  Here we let the 
stellar mass to the light ratio in $3.6\mu m$ as the free parameter to fit the observation with 
the prediction of different models.  
The goodness of fit for each galaxy is calculated by $\chi^2$ fit and the minimum value 
for this function provides the best $\Upsilon_{*}^{(3.6)}$ for each galaxy:
\begin{equation}
  \chi^2=\sum_{i=1}^{N} \frac{(V^{th}_{i}-V^{obs}_{i})^2}{{\sigma_{i}}^2},
\end{equation}
here $\sigma_i$ is the uncertainty of $V^{obs}_{i}$ and $N$ is the
number of data points. To see if MOG, NLG and MOND can fit the rotation curve of dwarf galaxies we fix  $\alpha = 8.89\pm 0.34$	and $\mu = 0.042\pm0.004~kpc^{-1}$ for MOG,  $\alpha=10.94 \pm 2.56$ and $\mu=0.059 \pm 0.028 kpc^{-1}$ for NLG and $a_{0}=1.0\times10^{-10}\,\mathrm{m\,s}^{-2}$ for MOND and let $\Upsilon_{*}^{(3.6)}$ to changes as the free parameter. The best value of $\Upsilon_{*}^{(3.6)}$ for different galaxies are given in the table (\ref{tab1}) and the average value for the best $\overline{\chi^2}$ for these gravity models are: $\overline{\chi^2}_{MOG}=1.14$,	$\overline{\chi^2}_{NLG}=0.84$ and $\overline{\chi^2}_{MOND}=1.99$ which are reasonable values for the $\chi^2$ fitting. }

{
The best value for the stellar mass to light ratio of dwarf galaxies are also given in table (\ref{tab1}). 
We note that the best value for $\Upsilon_{*}^{(3.6)}$ is different in various gravity models. This parameter for 
the case of MOND has the average value of $\bar{\Upsilon}_{*}^{(3.6)}(MOND)= 1.91$ and for MOG and NLG the 
corresponding values are $\bar{\Upsilon}_{*}^{(3.6)}(MOG)= 11.37$  and $\bar{\Upsilon}_{*}^{(3.6)}(NLG)= 6.91$ .
A large difference in the values of stellar mass to the light ratio in the dwarf galaxies is crucial point that 
can falsify the gravity models. A larger value of the stellar mass to light ratio of dwarf galaxies are also 
suggested in the other works as \citet{Masstolight-dwarf} and \cite{haghi} which is larger than the standard model
for the star formation models. Larger ${\Upsilon}_{*}$ for dwarf galaxies might results from the different history of
star formation in the dwarf galaxies than Milky Way. Having a diffuse medium as the dwarf galaxies may produce 
small mass stars. Since the luminosity of stars depends on the mass as $L\propto M^{\alpha}$ \citep{Kuiper,salaris}, population of 
small mass stars results in higher stellar mass to the light radio. On the other hand 
having a larger mass budget for the heavy mass stars produces stellar remnant 
and the result is a larger mass to the light ratio. This question might be resolved with the future 
observations of stellar populations in the dwarf galaxies.}

\setcounter{figure}{0}
\begin{figure*}
	\begin{center}
		\begin{tabular}{ccc}
			\includegraphics[width=55mm]{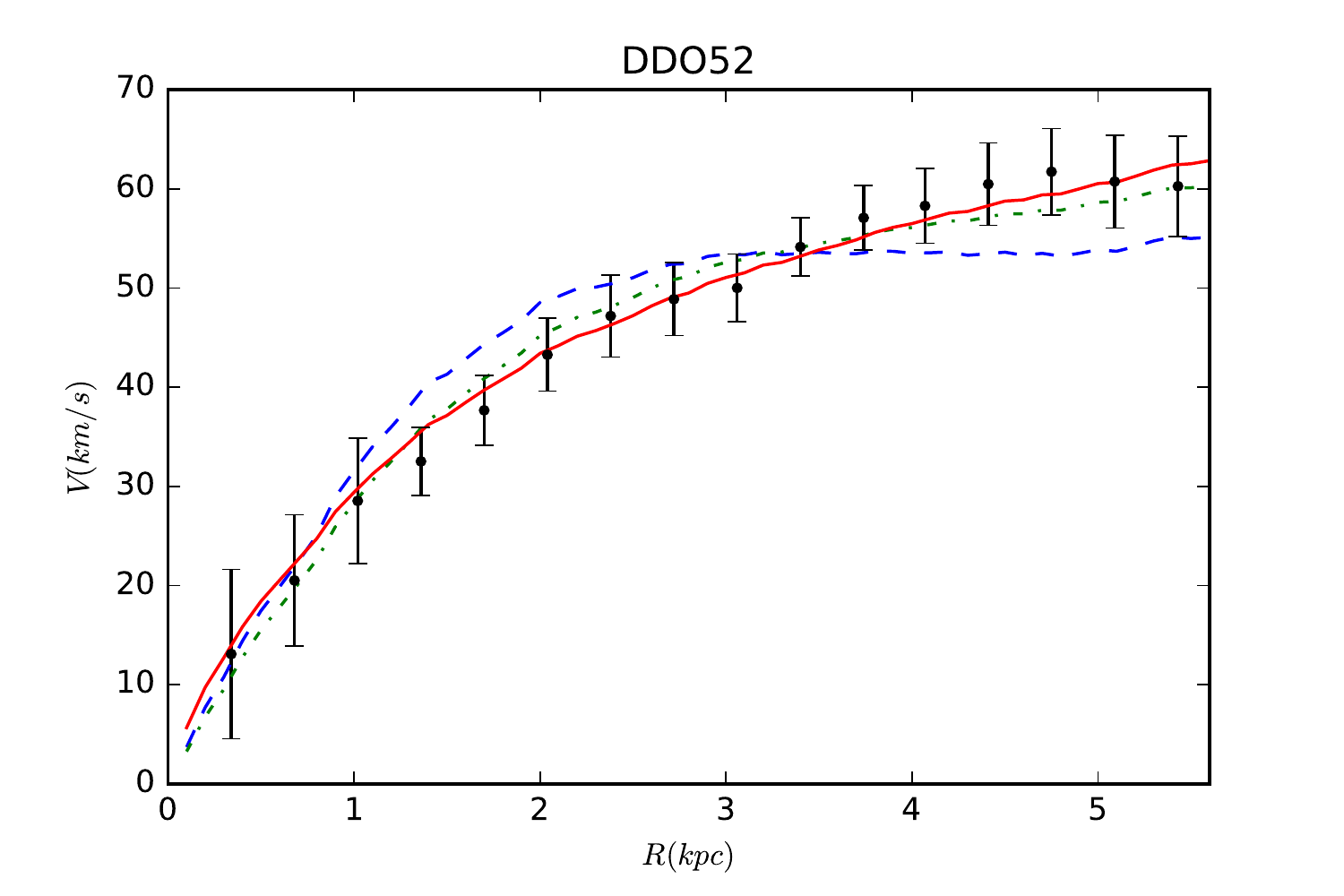}&
			\includegraphics[width=55mm]{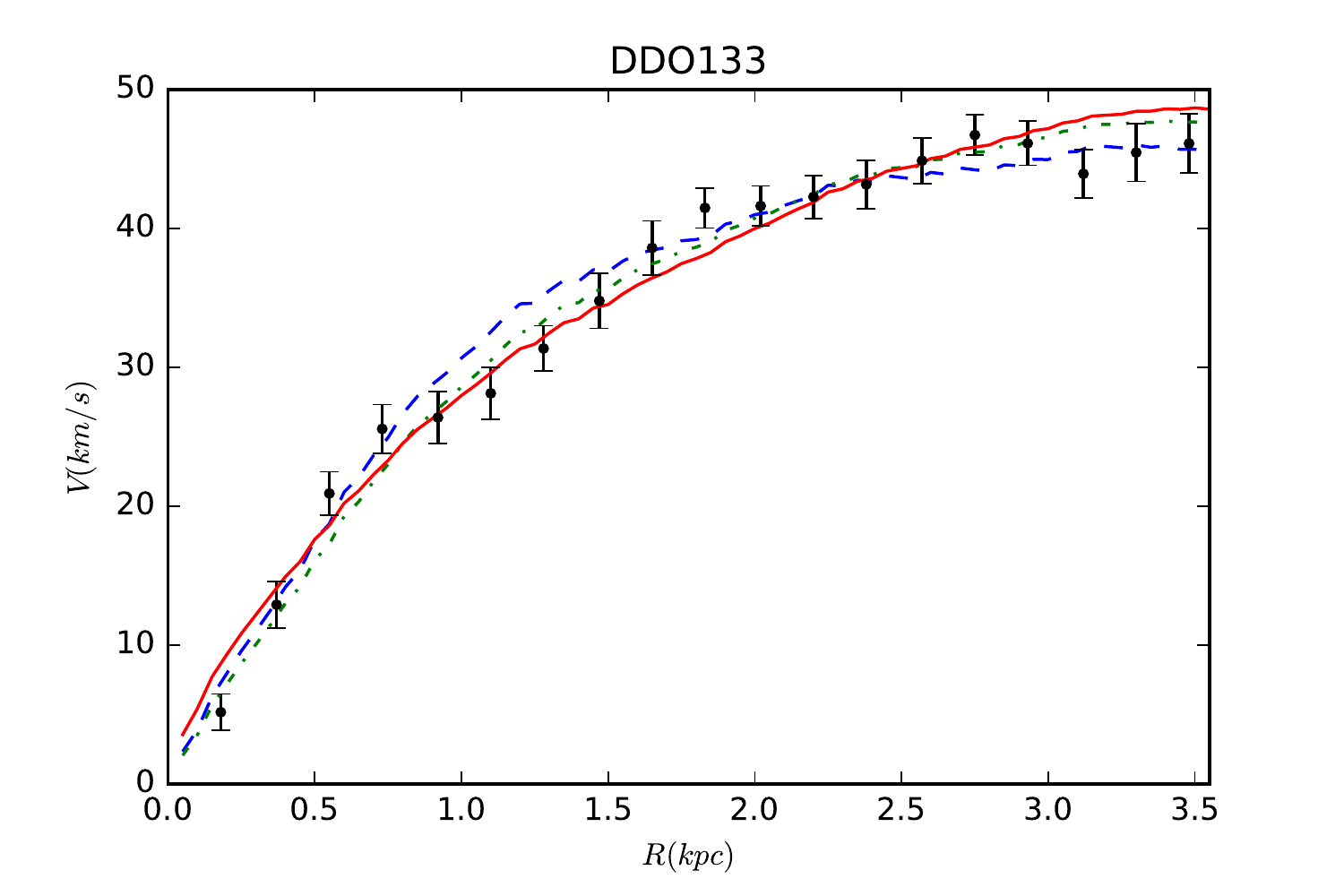}&
			\includegraphics[width=55mm]{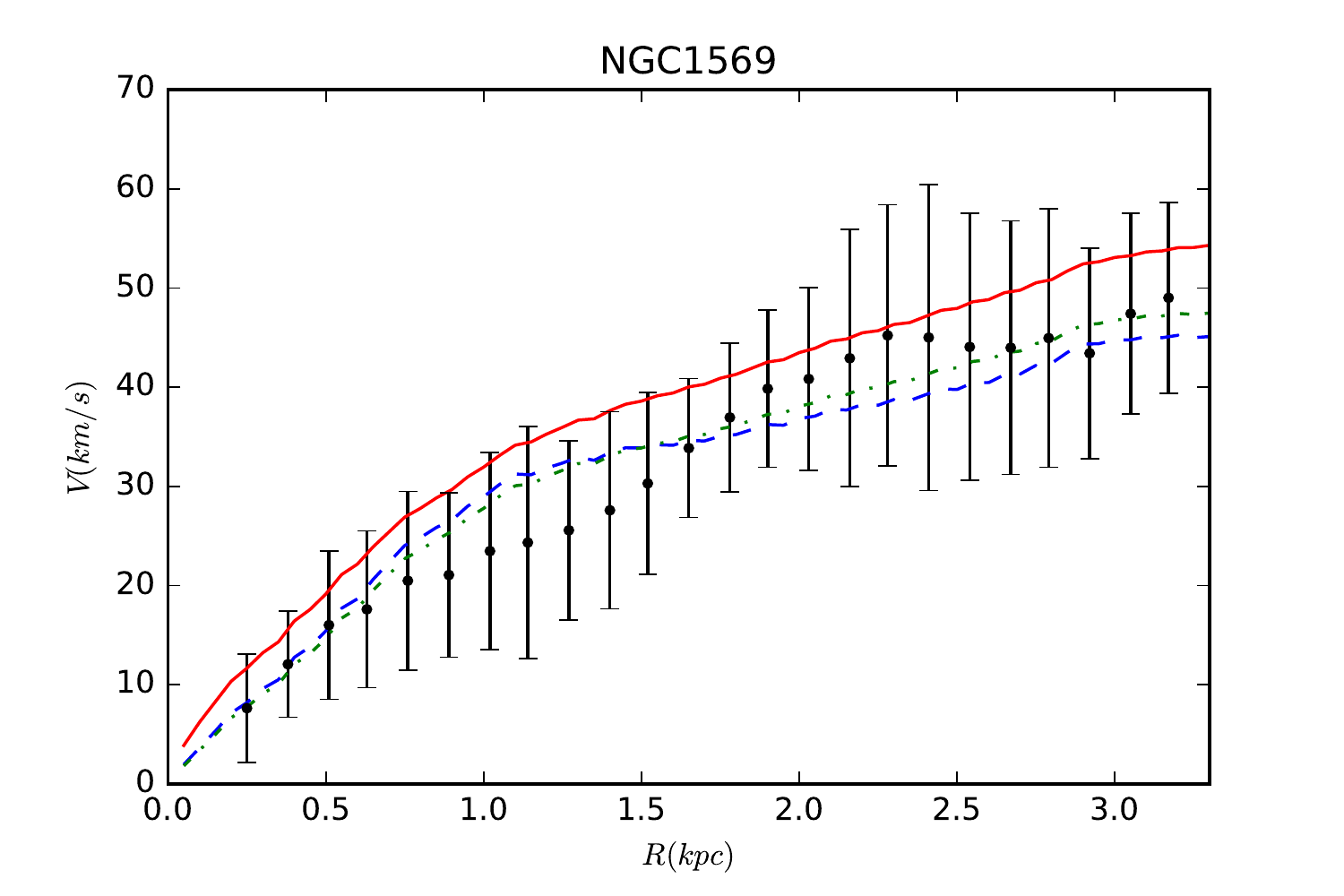}\\
			\includegraphics[width=55mm]{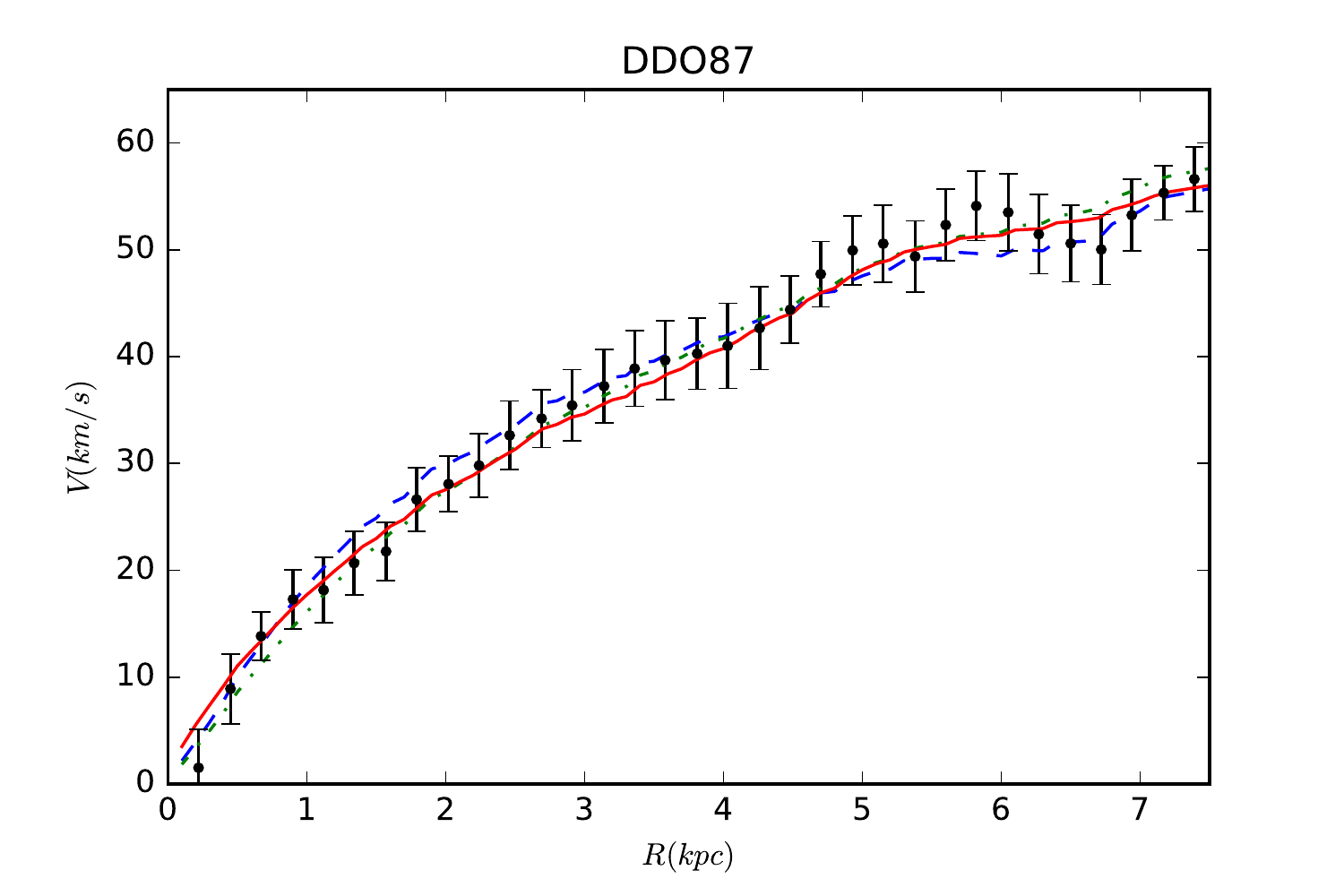}&
			\includegraphics[width=55mm]{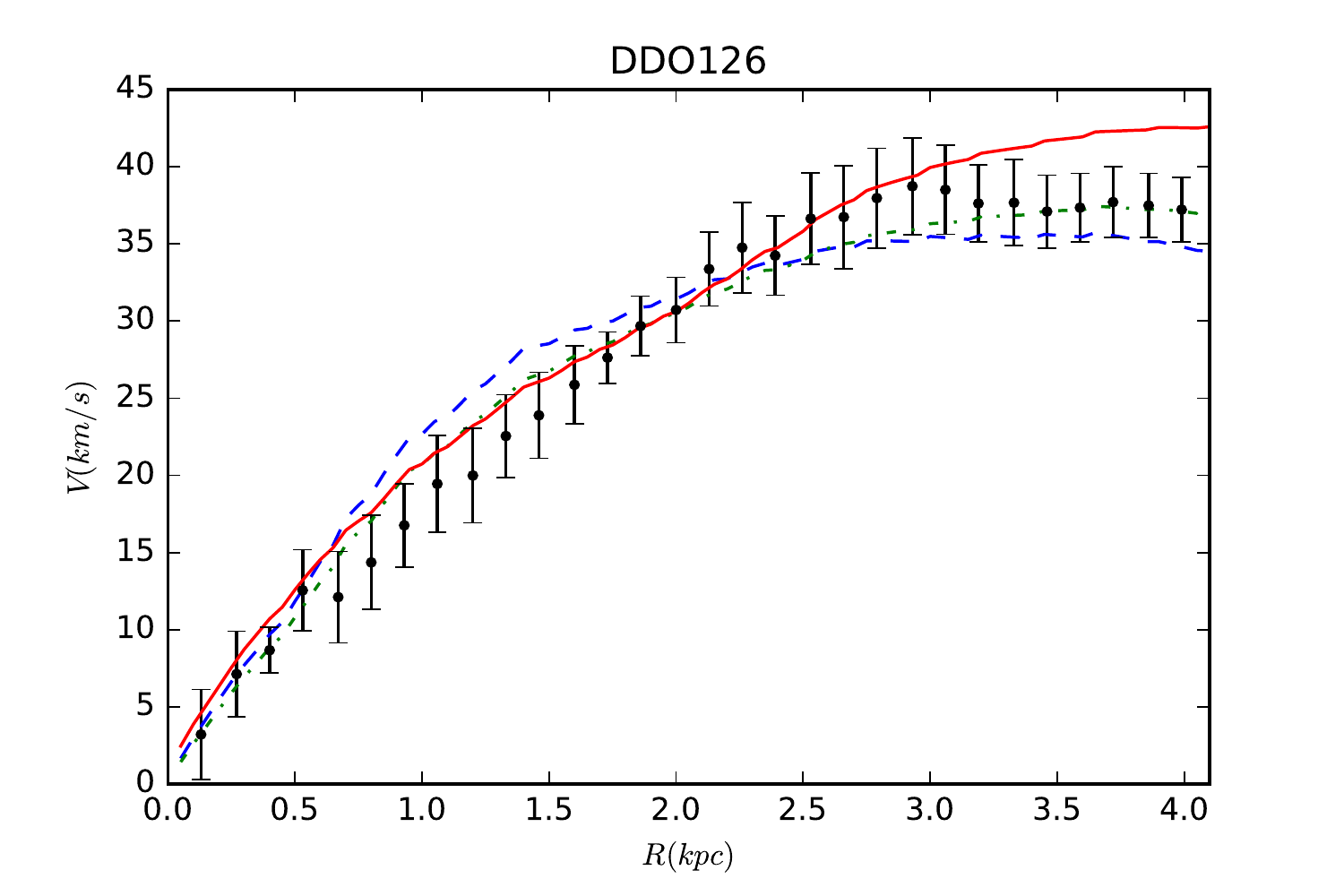}&
			\includegraphics[width=55mm]{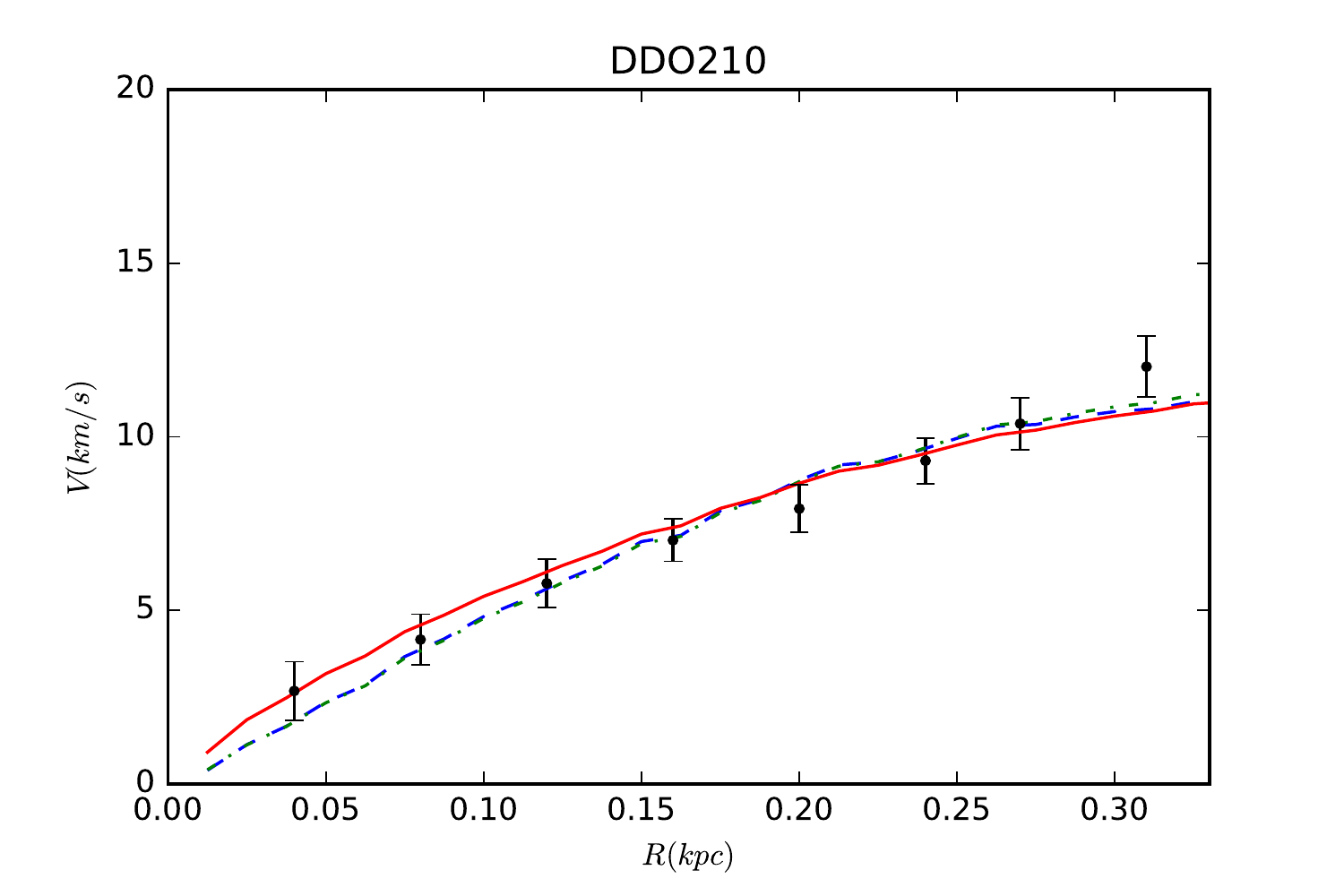}\\
			\includegraphics[width=55mm]{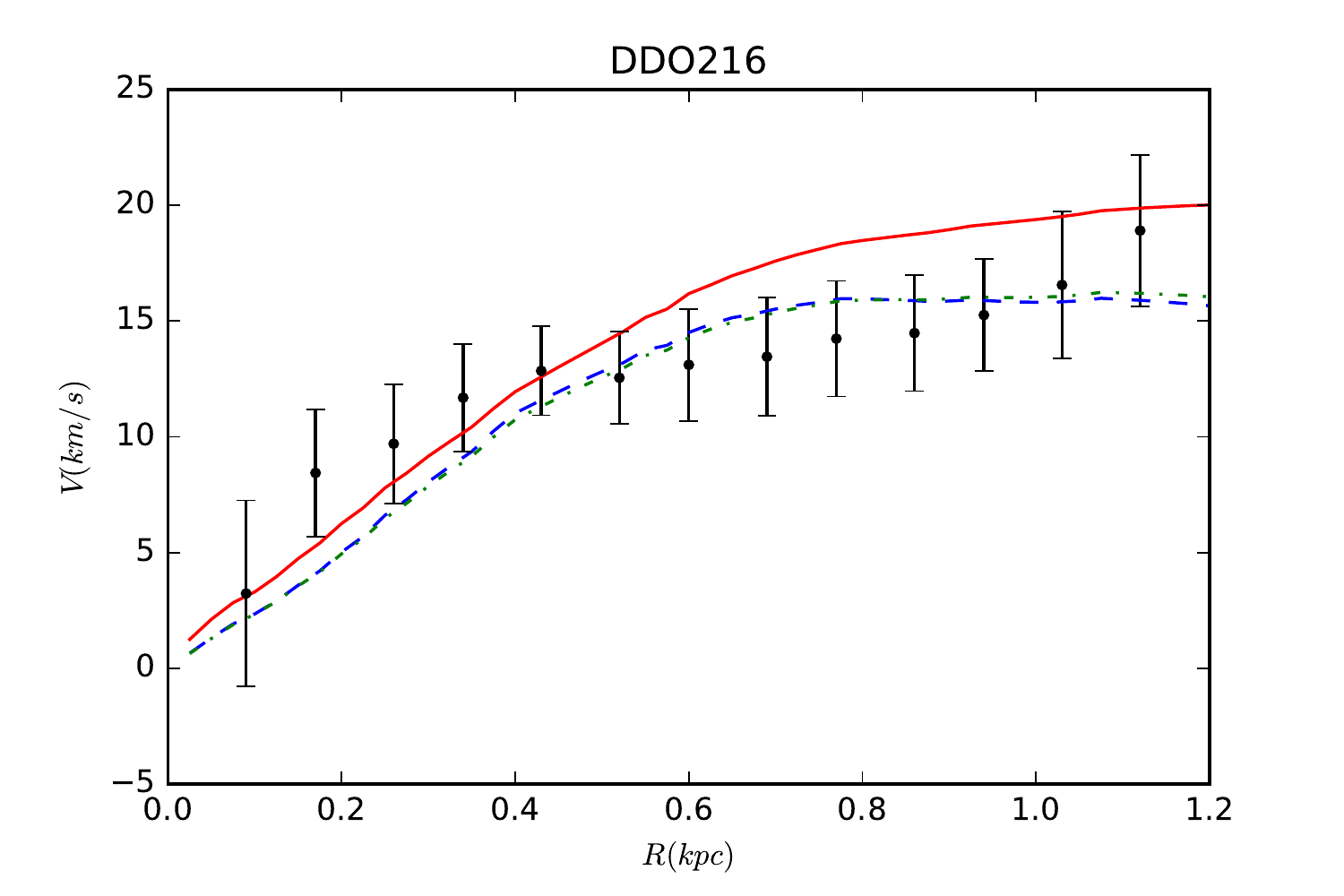} &
			\includegraphics[width=55mm]{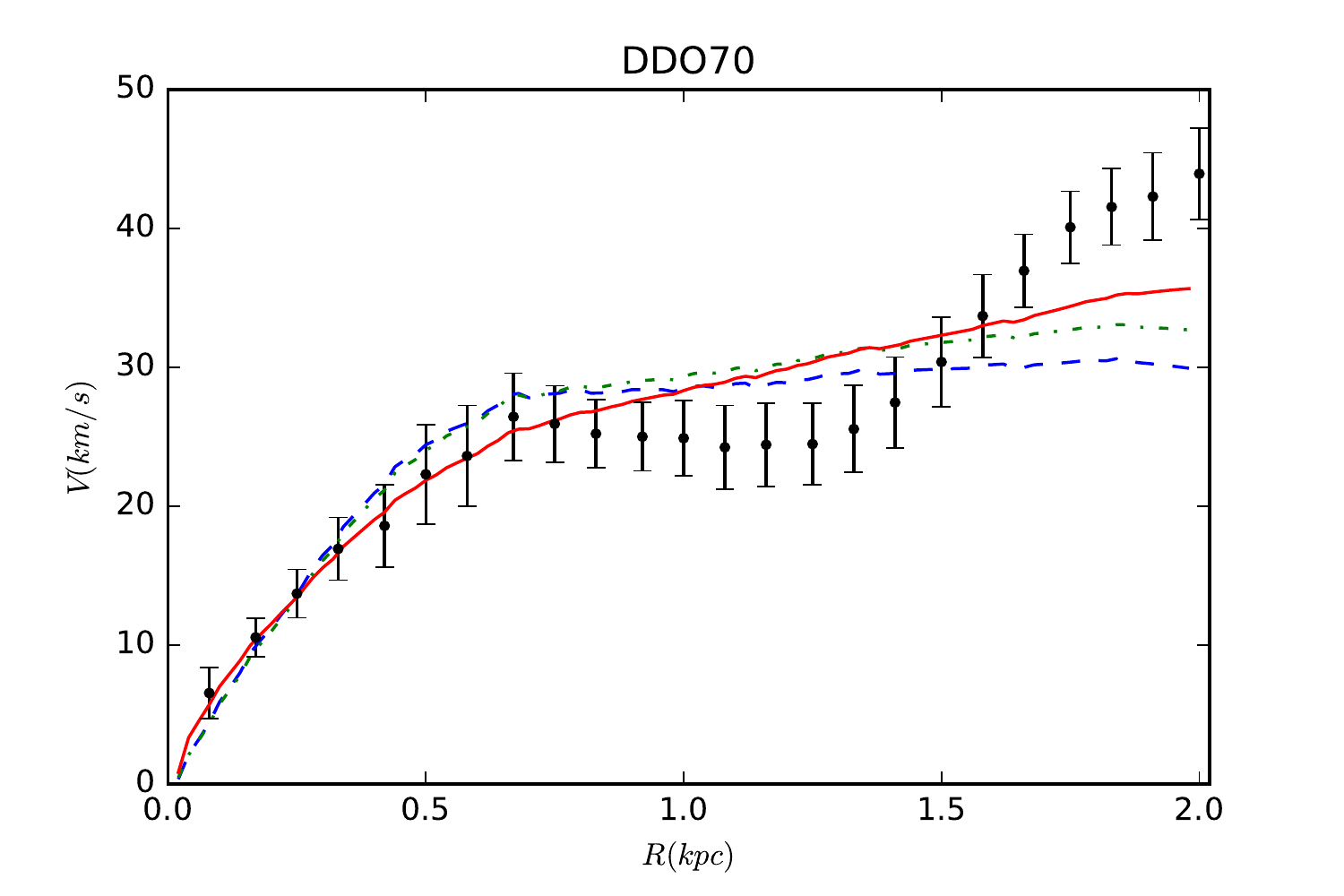} &
			\includegraphics[width=55mm]{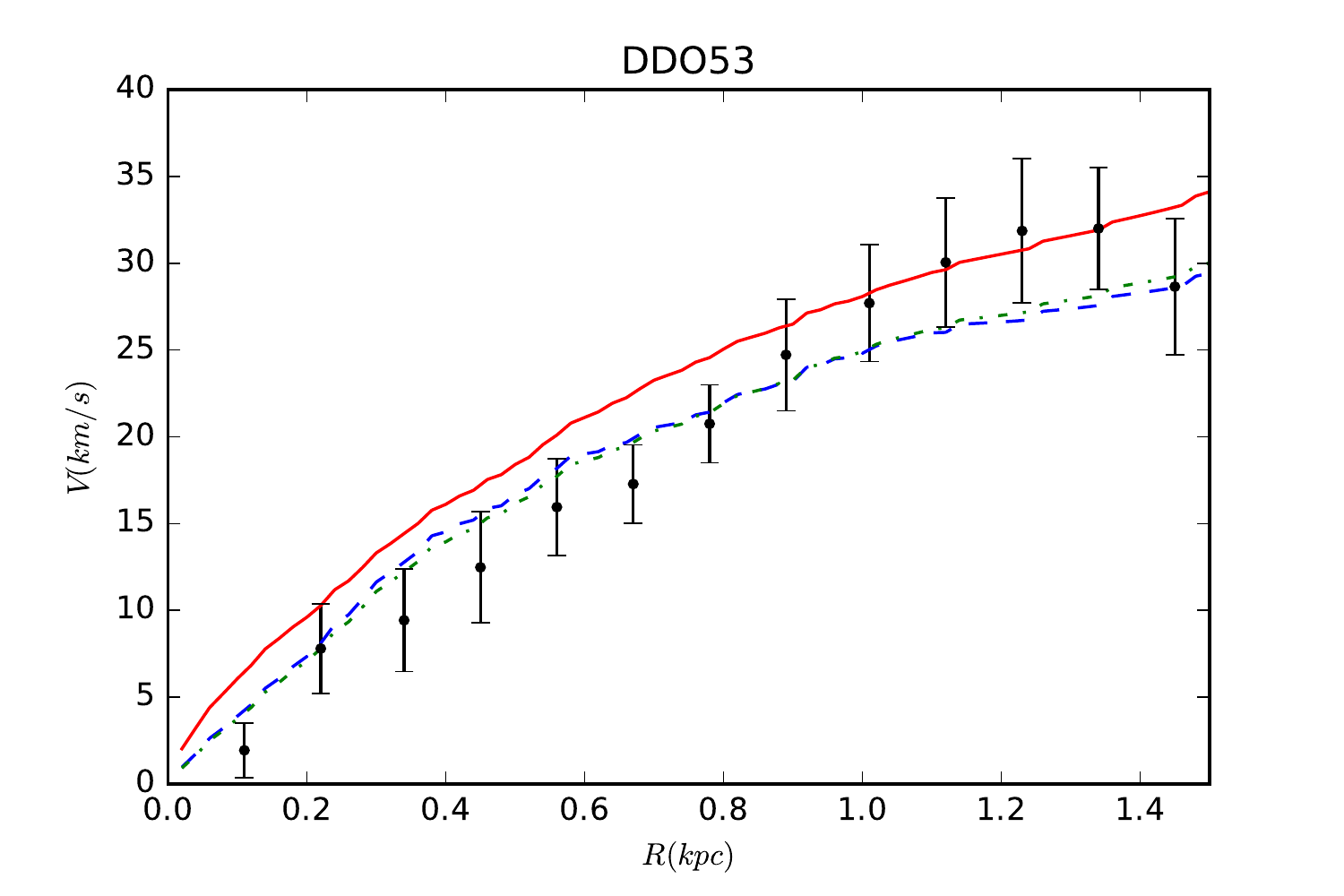}\\
		
			\includegraphics[width=55mm]{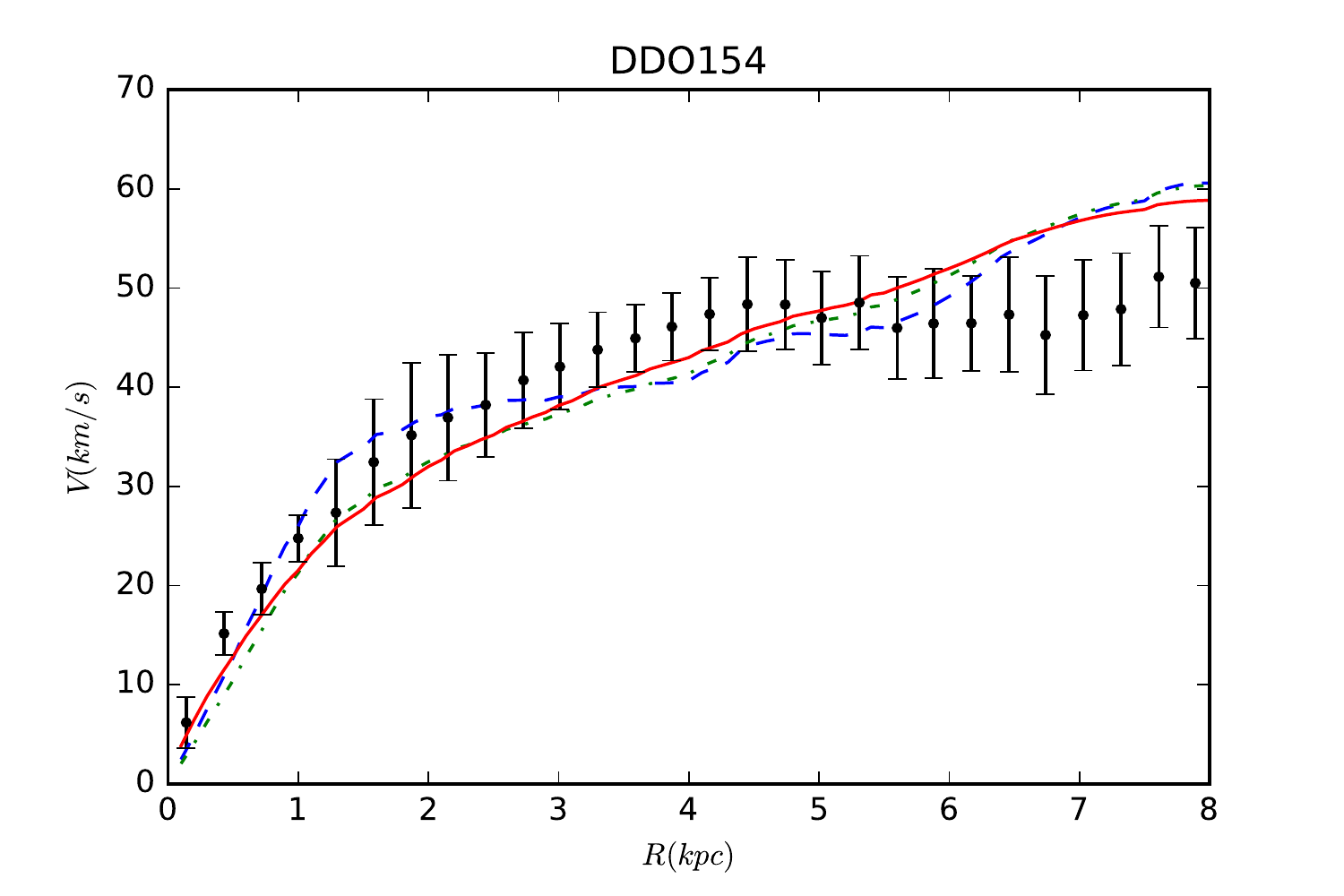}&
			\includegraphics[width=55mm]{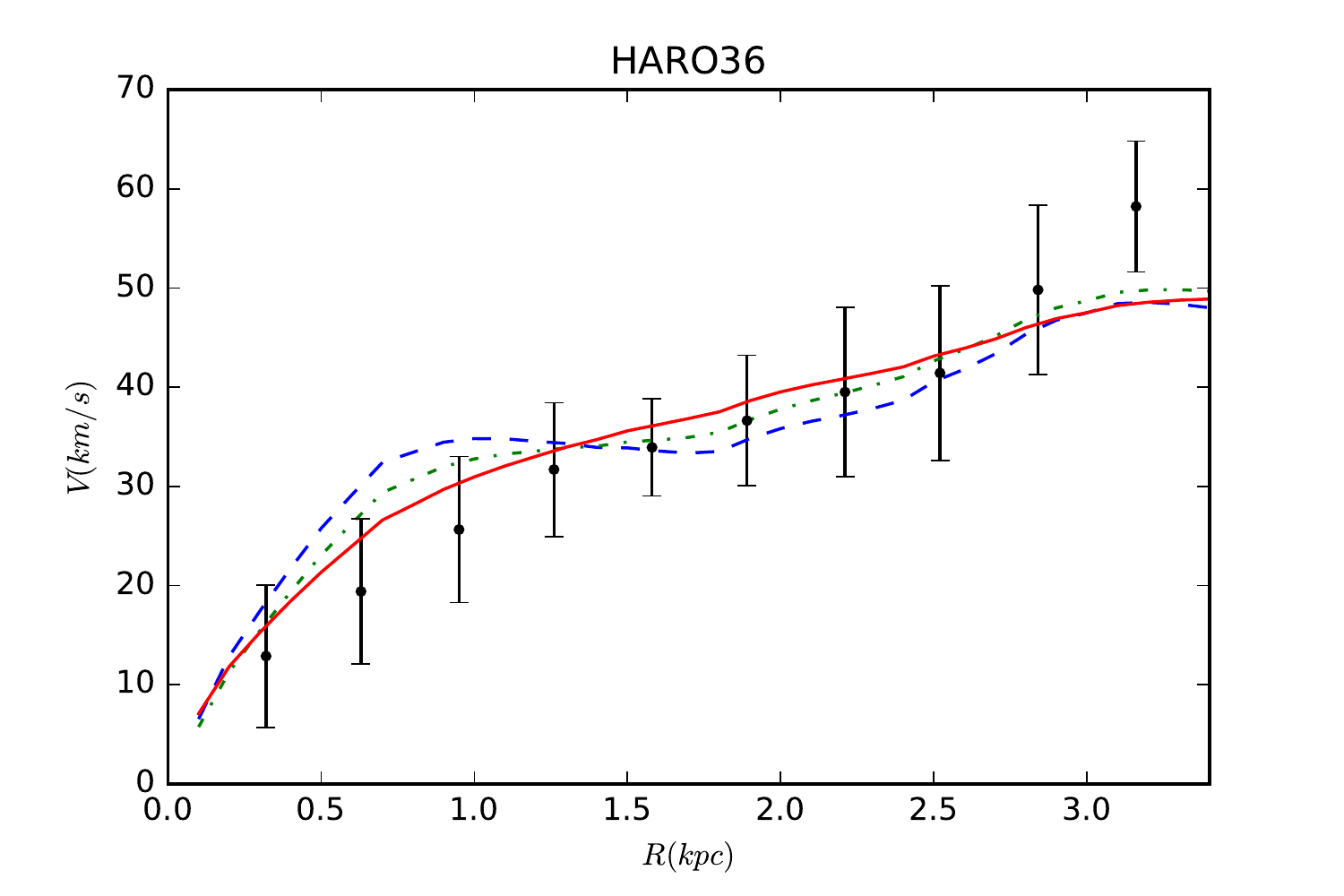}&
			\includegraphics[width=55mm]{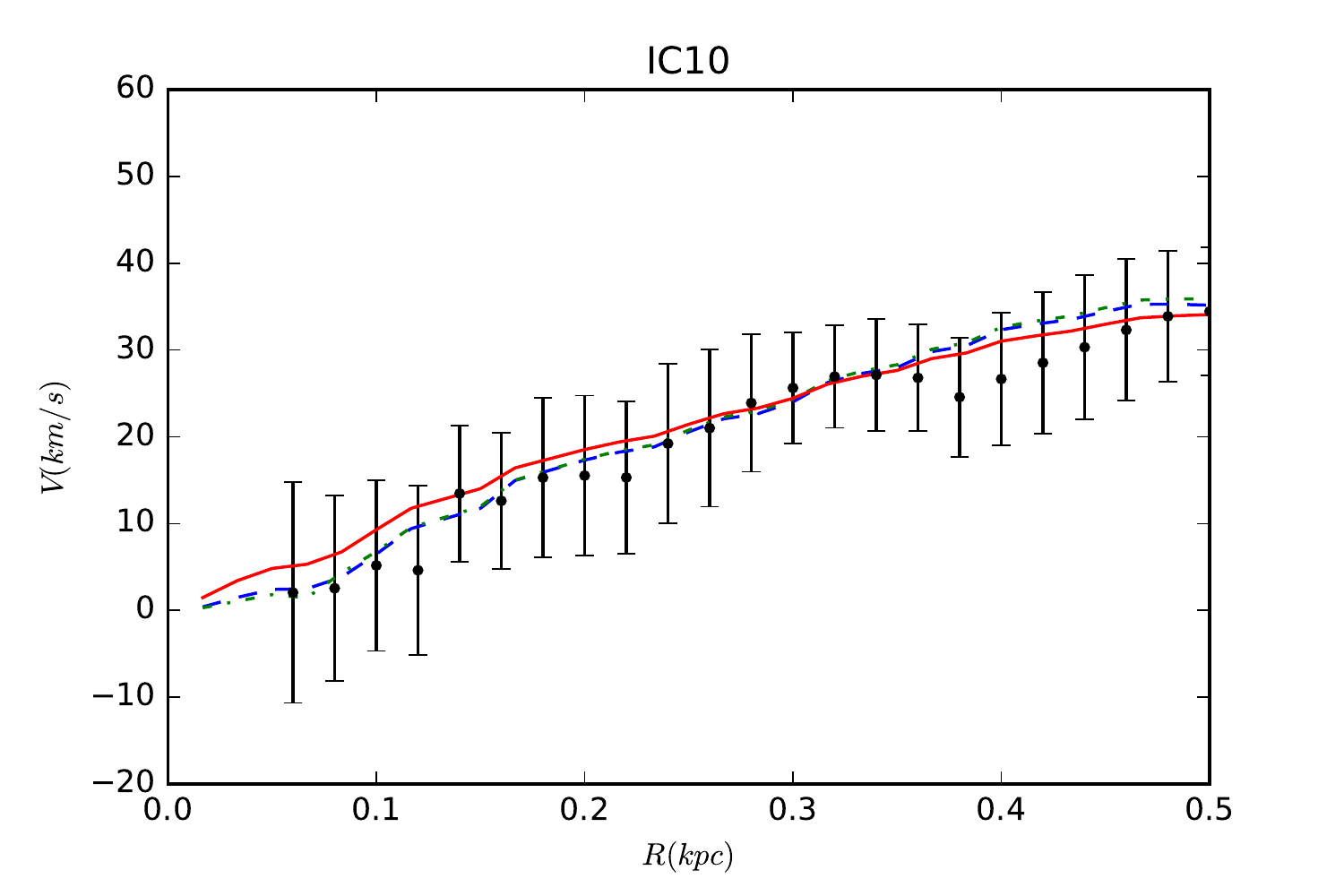}\\
			
			\includegraphics[width=55mm]{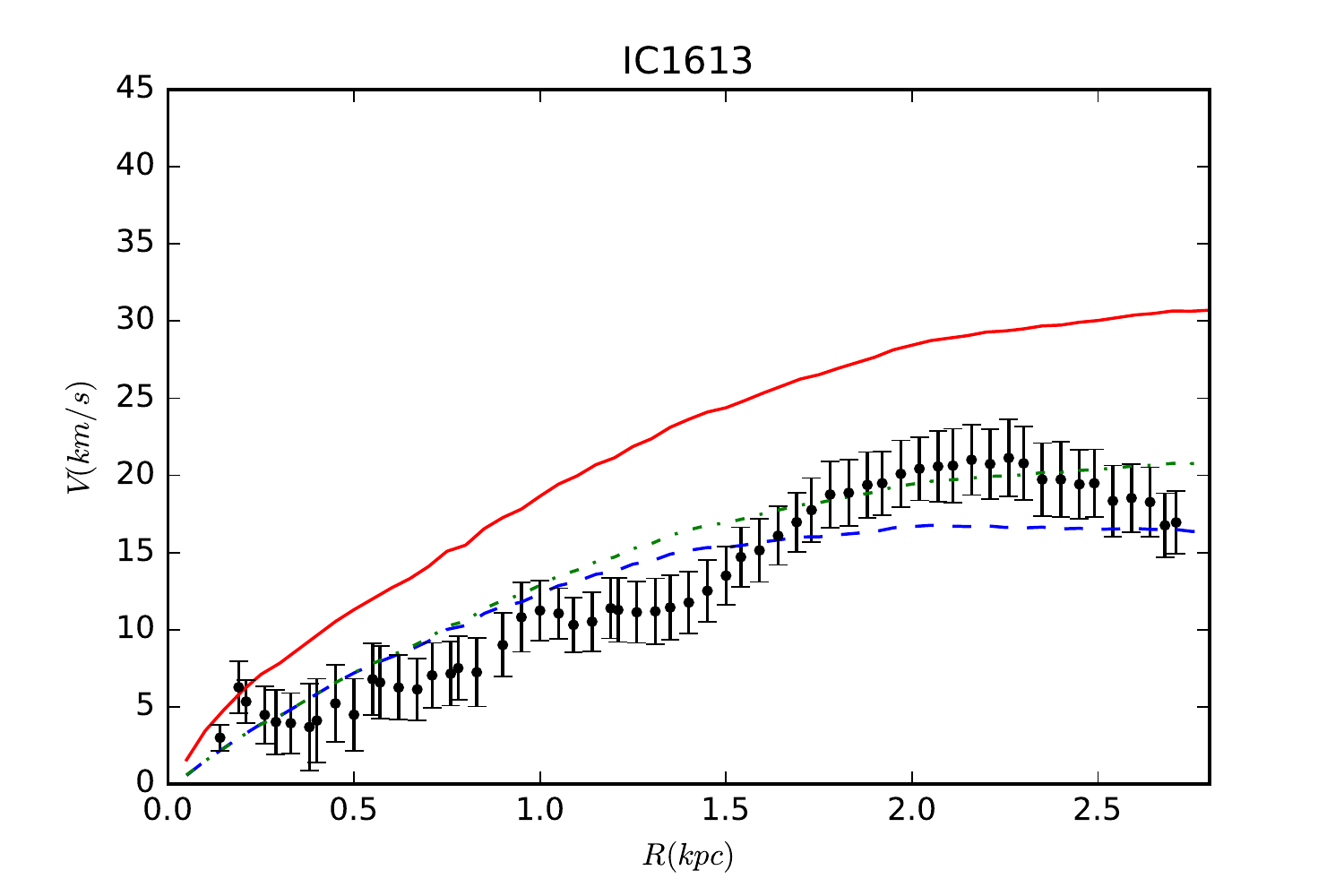}&
			\includegraphics[width=55mm]{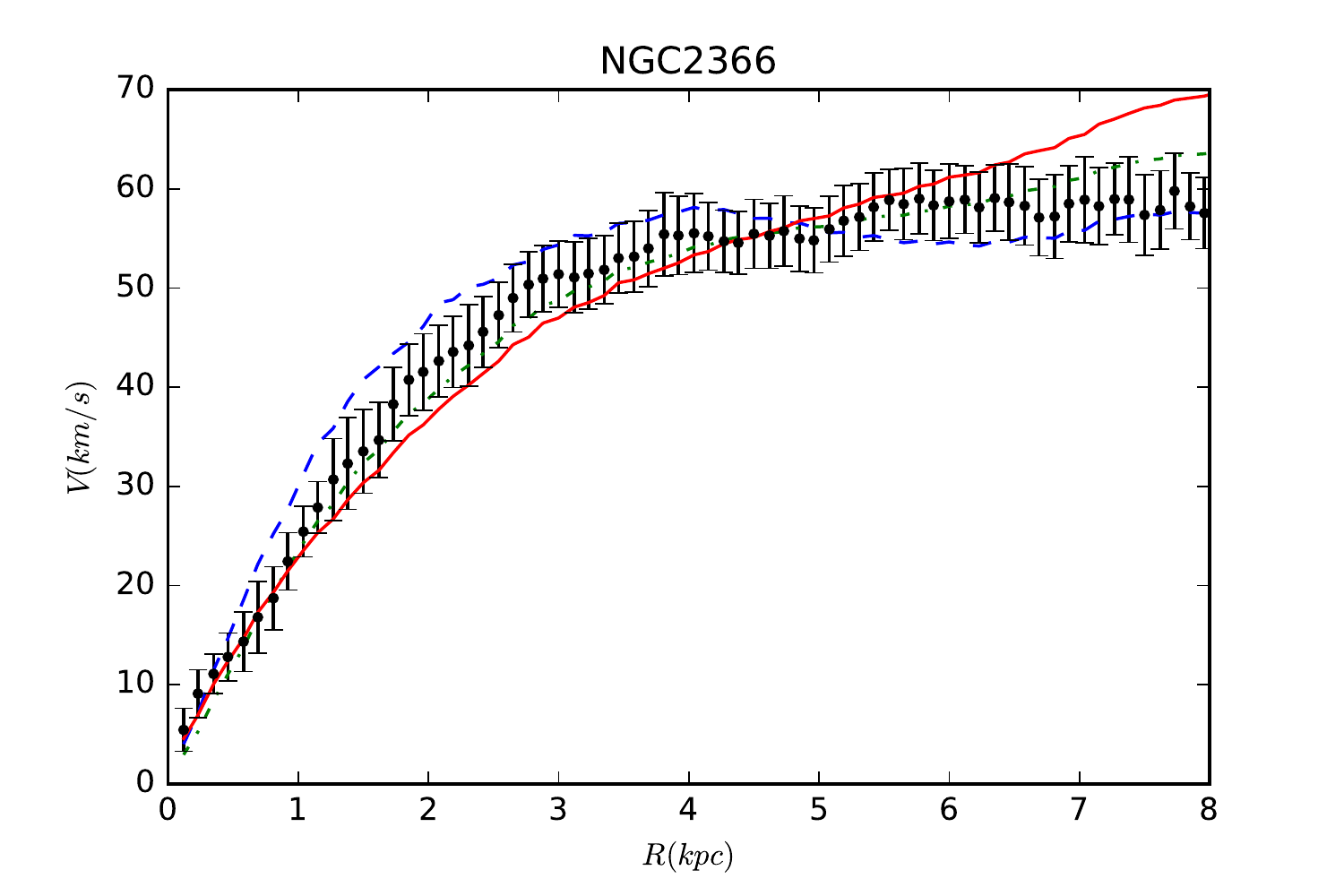}&
			\includegraphics[width=55mm]{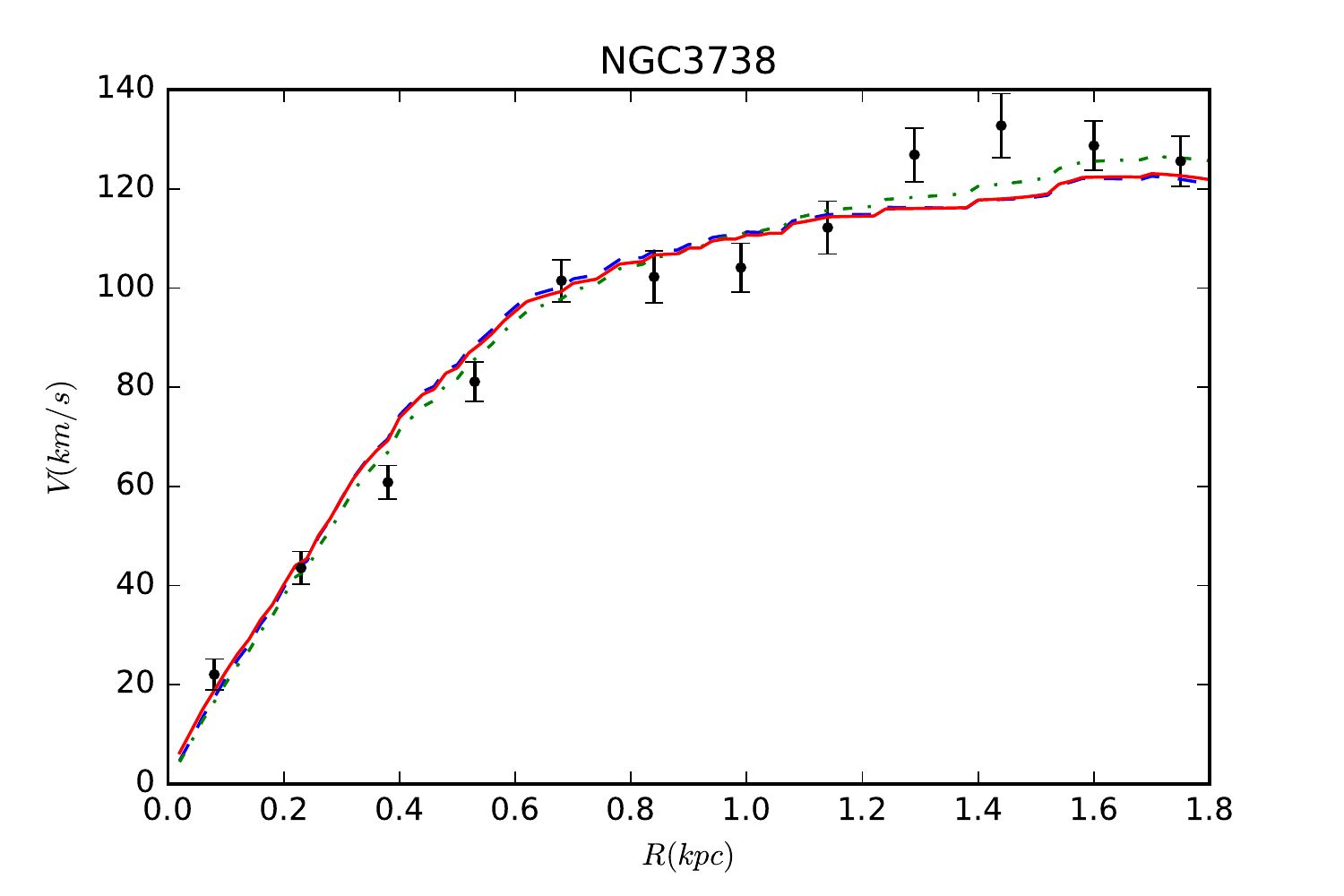}\\
			\includegraphics[width=55mm]{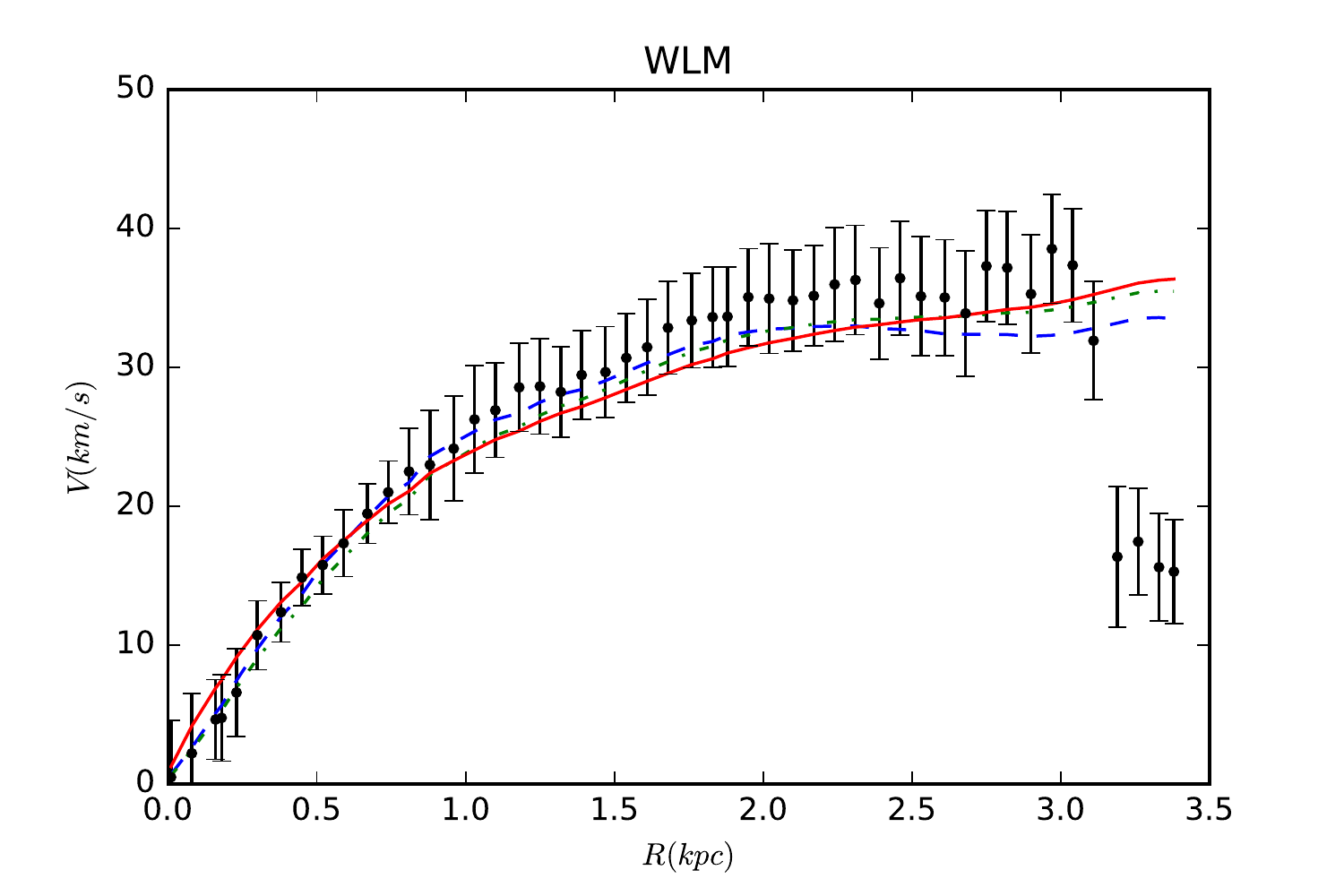}\\
		\end{tabular}
	\end{center}
	\caption {{\bf The best fits to the rotation velocity curves of the
          LITTLE THINGS sample. These curves are plotted using the
          universal value of $\alpha = 8.89\pm 0.34$	and $\mu = 0.042\pm0.004~kpc^{-1}$  (dashed line), $\alpha=10.94 \pm 2.56$ and $\mu=0.059 \pm 0.028 kpc^{-1}$ (dashed-dotted line), and the rotation curve for MOND
          using the best value of $a_{0}=1.0\times10^{-10}\,\mathrm{m\,s}^{-2}$ (solid line).
          }\label{fig1} }
	
\end{figure*}



\setcounter{figure}{1}
\begin{figure*}
\begin{center}
\begin{tabular}{ccc}
\end{tabular}
\end{center}
\end{figure*}


\begin{table*}
\begin{center}
  \caption[]{ {\bf Sub-set of dwarf galaxies from THINGS catalog
    \citep{dwarf}, with fits to Modified Gravity.  The columns are as
    follows: (1) name of the galaxy, (2) distance of the galaxy from
    us, (3) radius where the outermost part of the rotation curve is
    measured, (4) mass of gas, (5) stellar mass in $3.6~\mu$m band,
    (6) normalized $\chi^2$ for the best
        fit to the data in MOG, (7) normalized $\chi^2$ for the best
            fit to the data in NLG, (8) normalized $\chi^2$ for the best
    fit to the data in MOND, (9) best fit stellar mass to light ratio for MOG, (10) best fit stellar mass to light ratio for NLG, (11) best fit stellar
    mass to light ratio for MOND.} \label{tab1} }
\begin{tabular}{|c|c|c|c|c|c|c|c|c|c|c|c|}
\hline\hline
        & Distance & $R_{max}$   & $M_{gas}$         &$M_{star}$ & $\chi^2/N_{d.o.f}$ &$\chi^2/N_{d.o.f}$    &  $\chi^2/N_{d.o.f}$    & $\Upsilon^{(3.6)}_{\star MOG}$ &$\Upsilon^{(3.6)}_{\star NLG}$& $\Upsilon^{(3.6)}_{\star MOND}$ \\
Galaxy & (Mpc) &(kpc) & $(10^{10}M_\odot)$& $(10^{10}M_\odot)$& $(MOG)$ & $(NLG)$   & $(MOND)$  &    $M_\odot/L_\odot$   &    $M_\odot/L_\odot$ &$M_\odot/L_\odot$ \\ 
(1)   & (2)  & (3) & (4)  & (5) & (6) & (7) & (8) &(9) &(10) &(11)   \\
\hline
DDO 52& 10.3     & 5.43              &33.43               &  7.20   & 1.5  & 0.31  & 0.16 & $12.3_{-0.6}^{+0.5}$  & $6.5_{-0.4}^{+0.3}$ & $2.7_{-0.26}^{+0.28}$ \\
DDO 53  & 3.6     &1.45            &7.00               &   0.96         & 0.86 & 0.69   &  1.96    &  $5.9_{-0.7}^{+0.7}$  & $4.1_{-0.6}^{+0.7}$ & $0.2_{-0.08}^{+0.08}$  \\
DDO 70  &1.3      &2.00             & 3.80              &  1.24 & 3.5     & 2.8       & 1.66      & $7.8_{-0.40}^{+0.45}$   & $5.9_{-0.4}^{+0.2}$ & $1.42_{-0.2}^{+0.17}$ \\
DDO 87  &7.7     & 7.39              &29.12             &   6.18  & 0.37   & 0.26     & 0.23     & $15.2_{-0.45}^{+0.60}$  & $6.8_{-0.25}^{+0.28}$  &$1.2_{-0.1}^{+0.3}$ \\
DDO 126 &4.9      &3.99               & 16.36              &   2.27  & 1.1  & 0.38  & 1.24      & $8.6_{-0.4}^{+0.3}$  & $4.6_{-0.20}^{+0.24}$ & $0.1_{-0.04}^{+0.04}$ \\

DDO 133  &3.5     &3.48           & 12.85                 &   2.62 & 1.37  & 1.1   &  1.49   & $12.6_{-0.40}^{+0.25}$  & $7.6_{-0.3}^{+0.2}$ &$2_{-0.15}^{+0.15}$ \\
DDO 154  & 3.7      &2.59             & 35.27              &  1.31 &  1.25    & 1.31        & 0.96     & $42.5_{-2.0}^{+2.1}$  & $19_{-1.0}^{+1.2}$ &$3.2_{-0.65}^{+0.70}$ \\
DDO 210  & 0.9      &0.31             & 0.14              &  0.04 & 0.58     & 0.51        & 0.46      & $27.6_{-1.9}^{+2.0}$   & $24_{-1.7}^{+1.8}$ & $3.3_{-0.54}^{+0.35}$ \\
DDO 216  &1.1      &1.12             & 0.49              &  1.60  & 0.6 & 0.61           & 1.18    & $3.2_{-0.4}^{+0.4}$   & $2.6_{-0.3}^{+0.35}$ & $0.3_{-0.12}^{+0.12}$\\
Haro36  &9.3      &3.16             & 11.16            &  5.81 & 0.62     &  0.37       & 0.35      & $8.0_{-1.1}^{+1.2}$   & $5.1_{-1.2}^{+1.25}$ & $1.4_{-0.5}^{+0.51}$\\
IC 10  &0.7    &0.54              & 1.65               &   11.81  & 0.07   & 0.07        & 0.09    & $0.9_{-0.14}^{+0.12}$  & $0.8_{-0.18}^{+0.08}$ &$0.5_{-0.12}^{+0.10}$ \\
IC 1613  &0.7    &2.71              & 5.93               &  1.94  & 1.48   & 0.98        & 17.4   & $1.9_{-0.13}^{+0.22}$   & $1_{-0.16}^{+0.05}$ & $0.1_{-0.05}^{+0.05}$  \\

NGC 1569 &3.4     &3.05           & 20.24               &   20.69 & 0.16  & 0.1         & 0.45    & $1.89_{-0.3}^{+0.25}$   & $1.1_{-0.20}^{+0.20}$ & $0.5_{-0.038}^{+0.038}$  \\
NGC2366  &3.4      &8.08            & 108.24              &  10.81 & 2.1     & 1.35        & 1.8     & $7.8_{-0.25}^{+0.13}$  & $2.8_{-0.07}^{+0.15}$ & $0.3_{-0.05}^{+0.15}$ \\
NGC3738  &4.9      &1.75             & 12.58             &  12.48 & 0.92    & 0.49        & 1.69     & $11.6_{-0.35}^{+0.37}$  & $9_{-0.25}^{+0.30}$ & $11.2_{-0.30}^{+0.35}$ \\
WLM  &1.0    &3.04            & 7.96              &  1.23 & 1.8     & 2.14        & 2.37     & $14.2_{-0.5}^{+0.6}$  & $9.6_{-0.35}^{+0.38}$ & $2.2_{-0.25}^{+0.15}$ \\


\hline
\end{tabular}
\end{center}
\end{table*}

\section{Conclusions}
 {To test MOG, Non-Local Gravity and MOND models as the alternative models 
 for the dark matter, we compared the theoretical
 rotation curves predicted by these three gravity models with the observed
 data for sixteen dwarf galaxies in the LITTLE THINGS catalog. 
 The gravitational acceleration due to a point source in the weak
 field limit of MOG and Non-Local Gravity involves
 two parameters: $\alpha$ determines the gravitational coupling
 strength via $G=G_N(1+\alpha)$ and $\mu$ as the inverse of the
 characteristic length of the repulsive Yukawa force.  At distances
 much greater than $\mu^{-1}$, the repulsive term is negligible.  For
 each of MOG and Non-Local Gravity, we fix the parameters of 
 $\mu$ and $\alpha$ with the universal values that has been reported in 
 \citep{rahvar1,rahvar3} and
  analyze the rotation curve of dwarf galaxies, using $\Upsilon_{*}$ at $3.6\mu m$ as the only free 
parameter of model. The same procedure for MOND has been done like the other gravity models. We fixed $a_{0}=1.0\times10^{-10}\,\mathrm{m\,s}^{-2}$ as the universal value which is consistent with the value obtained previously by fitting to spiral galaxies \citep{sanders and McG}. }
  
  { For the two modified gravity models of MOG and NLG, the value of $\Upsilon_{*}$ in the dwarf galaxies is larger than the conventional value in the Milky Way while for MOND we obtain a compatible value with the spiral galaxies. The stellar mass 
  to the light ratio is a function of stellar population inside a galaxy where for population with larger mass 
  stars this parameter would be smaller and for small mass population of stars, that will be larger. For diffused 
  mediums as the dwarf galaxies, the history of star formation might be different and produce small mass stars. The result would be a larger stellar mass to the light ratio. On the other hand
  the remnant of heavy stars also can produce a larger stellar mass to the light ratio. This phenomenon can be examined by direct 
  observations of stellar populations in the dwarf galaxies. Future telescope may resolve stars in the nearby dwarf 
  galaxies and can rule out either MOG/NLG models or MOND. }



\label{conc}
\section*{Acknowledgments}
{\bf We thank John Moffat and Martin Green for their useful comments and improving the text of 
paper. Also we thank referee for his/her useful comments for improving this work. }

\label{lastpage}


\begin{thebibliography}{}

\bibitem[Akerib et.al(2014)]{ake} Akerib, D. S., Araujo, H. M., Bai,
  X., et al. 2014, Physical Review Letters, 112, 091303

\bibitem[Aldrovandi \& Pereira(2013)]{tp2} Aldrovandi, R., \& Pereira,
  J.~G.\ 2013, Teleparallel Gravity: An Introduction, Fundamental
  Theories of Physics, Volume 173.~ISBN 978-94-007-5142-2.~Springer
  Science+Business Media Dordrecht, 2013,

\bibitem[Angloher et.al(2012)]{ang} Angloher, G., Bauer, M., Bavykina,
  I., et al. 2012, European Physical Journal C, 72, 1971

\bibitem[Angus et al.(2007)]{MOND-cluster2} Angus, G.~W., Shan, H.~Y., Zhao, H.~S., \& Famaey, B.\ 2007, \apjl, 654, L13 


\bibitem[Bekenstein(2004)]{beken} Bekenstein, J.~D.\ 2004, Phys Rev D,
  70, 083509

\bibitem[Bevington \& Robinson(2003)]{confidence} Bevington, P.~R., \&
  Robinson, D.~K.\ 2003, Data reduction and error analysis for the
  physical sciences, 3rd ed., by Philip R.~Bevington, and Keith
  D.~Robinson.~Boston, MA: McGraw-Hill, ISBN 0-07-247227-8, 2003.,

\bibitem[Blagojevi{\'c} \& Hehl(2012)] {tp1} Blagojevi{\'c}, M., \&
  Hehl, F.~W.\ 2012, arXiv:1210.3775

\bibitem[Blumenthal et al.(1984)]{StFormation} Blumenthal, G.~R., Faber, S.~M., Primack, J.~R., \& Rees, M.~J.\ 1984, \nat, 311, 517 

\bibitem[Bruneton et al.(2009)]{hybrid1} Bruneton, J.-P., Liberati, S., Sindoni, L., \& Famaey, B.\ 2009, \jcap, 3, 021 


\bibitem[de Blok \& McGaugh(1997)]{DM_in_dwarf} de Blok, W.~J.~G., \&
  McGaugh, S.~S.\ 1997, 290, 533

\bibitem[Eisenstein et al.(2005)]{BAO} Eisenstein, D.~J., Zehavi, I., Hogg, D.~W., et al.\ 2005, \apj, 633, 560 

\bibitem[Gaitskell(2004)]{gal} Gaitskell, R. J. 2004, Annual Review of
  Nuclear and Particle Science, 54, 315

\bibitem[Haghi \& Amiri(2016)]{haghi} Haghi, H., \& Amiri, V.\ 2016, \mnras, 463, 1944 


\bibitem[Hehl \& Mashhoon(2009)]{mashhoon} Hehl, F.~W., \& Mashhoon,
  B.\ 2009, Physics Letters B, 673, 279

\bibitem[Hunter et al.(2012)]{Lthings} Hunter, D.~A., Ficut-Vicas, D.,
  Ashley, T., et al.\ 2012, 144, 134

\bibitem[Hunter et.al(2015)]{dwarf} Hunter, A. et.al
  2015,arXiv:1502.01281v1 [astro-ph.GA] 4 Feb 2015


\bibitem[Khoury(2015)]{hybrid2} Khoury, J.\ 2015, \prd, 91, 024022 

\bibitem[Kuiper(1938)]{Kuiper} Kuiper, G.~P.\ 1938, \apj, 88, 472 


\bibitem[Maluf(2013)]{tp3} Maluf, J. W. \ 2013, Ann Phys 525, 339.

\bibitem[Markevitch et al.(2004)]{bullet} Markevitch, M., Gonzalez,
  A.~H., Clowe, D., et al.\ 2004, ApJ, 606, 819

\bibitem[Mashhoon(1993)]{mashhoon93} Mashhoon, B.\ 1993, Phys Rev A,
  47, 4498

\bibitem[Mashhoon(2007)]{mashhoon2007} Mashhoon, B.\ 2007, Annalen der
  Physik, 519, 57

\bibitem[McGaugh et al.(2016)]{McGaugh} McGaugh, S., Lelli, F., \& Schombert, J.\ 2016, arXiv:1609.05917 

\bibitem[Milgrom(1983)]{milgrom} Milgrom, M.\ 1983, ApJ, 270, 365

\bibitem[Moffat(2006)]{moffat06} Moffat, J.~W.,2006, JCAP, 3, 4

\bibitem[Moffat \& Toth(2008)]{moffattoth} Moffat, J.~W., Toth, V.~T.\
  2008, ApJ, 680, 1158

\bibitem[Moffat \& Toth(2009)]{moffat09} Moffat, J. W., Toth, V. T.,
  2009, Classical and Quantum Gravity 26 (8), 085002

\bibitem[Moffat \& Rahvar(2013)]{rahvar1} Moffat, J.~W., \& Rahvar,
  S.\ 2013, MNRAS, 436, 1439

\bibitem[Moffat \& Rahvar(2014)]{rahvar2} Moffat, J.~W., \& Rahvar,
  S.\ 2014, MNRAS, 441, 3724

\bibitem[Moore et al.(2001)]{moore} Moore, B., Calcaneo-Roldan,
  C.,Stadel, J., et al. 2001, Phys. Rev. D, 64, 063508

\bibitem[Oh et al.(2015)]{Massmodel} Oh, S.-H., Hunter, D.~A., Brinks,
  E., et al.\ 2015, AJ, 149, 180

\bibitem[Rahvar \& Mashhoon(2014)]{rm} Rahvar, S., \& Mashhoon, B.\
  2014, Phys. Rev. D, 89, 104011

\bibitem[Rahvar \& Mashhoon(2014)]{rahvar3} Rahvar, S., \& Mashhoon,
  B.\ 2014, Phys Rev D, 89, 104011
 
\bibitem[Rubin et al.(1965)]{rubin1} Rubin, V.~C., Burbidge, E.~M.,
  Burbidge, G.~R., \& Prendergast, K.~H.\ 1965, ApJ, 141, 885

\bibitem[Rubin et al.(1970)]{rubin2} Rubin V.~C.  \& Ford, W.~K.,
  Jr. 1970, ApJ, 159, 379

\bibitem[Salaris \& Cassisi(2005)]{salaris} Salaris, M., \& Cassisi, S.\ 2005, Evolution of Stars and Stellar Populations, by Maurizio Salaris, Santi Cassisi, pp.~400.~ISBN 0-470-09220-3.~Wiley-VCH , December 2005., 400 


\bibitem[Sanders(1999)]{MONDcluster} Sanders, R.~H.\ 1999, \apjl, 512, L23 

\bibitem[Sanders \& McGaugh(2002)]{sanders and McG} Sanders, R.~H., \& McGaugh, S.~S.\ 2002, \araa, 40, 263 


\bibitem[Sanders \& Land(2008)]{Sanders} Sanders, R.~H., \& Land, D.~D.\ 2008, \mnras, 389, 701 

\bibitem[Sanders(2007)]{MOND-cluster1} Sanders, R.~H.\ 2007, \mnras, 380, 331 

\bibitem[Spergel(2015)]{CDM} Spergel, D.~N.\ 2015, Science, 347, 1100 



\bibitem[Swaters et al.(2011)]{Masstolight-dwarf} Swaters, R.~A., Sancisi, R., van Albada, T.~S., \& van der Hulst, J.~M.\ 2011, \apj, 729, 118 




\bibitem[Swaters et al.(2010)]{dwarf Mond} Swaters, R.~A., Sanders,
  R.~H., \& McGaugh, S.~S.\ 2010, ApJ, 718, 380

\bibitem[Zwicky(1937)]{zw} Zwicky, F. 1937, ApJ, 86, 217





\end{thebibliography}
\end{document}